\documentclass[fleqn,usenatbib]{mnras}

\usepackage{newtxtext,newtxmath}

\usepackage[T1]{fontenc}
\usepackage{ae,aecompl}

\usepackage{graphicx}	
\usepackage{amsmath}	
\usepackage{amssymb}	
\usepackage{physics}    
\usepackage{pdflscape}
\usepackage{mathtools}  

\newcommand{\mb}{\mathbf}

\newcommand{\mc}{\mathcal}

\DeclarePairedDelimiter{\nint}\lfloor\rceil

\DeclarePairedDelimiter{\ceil}{\lceil}{\rceil}
\DeclarePairedDelimiter{\floor}{\lfloor}{\rfloor}

\title[Revisiting the Fast Folding Algorithm]{Optimal periodicity searching: Revisiting the Fast Folding Algorithm for large-scale pulsar surveys}

\author[V. Morello et al.]{
V. Morello,$^{1}$\thanks{E-mail: vincent.morello@postgrad.manchester.ac.uk}
E.D. Barr,$^{2}$
B.W. Stappers,$^{1}$
E.F. Keane,$^{3,1}$
and A.G. Lyne$^{1}$
\\
$^{1}$Jodrell Bank Centre for Astrophysics, Department of Physics and Astronomy, The University of Manchester, Manchester M13 9PL \\
$^{2}$Max-Planck-Institut f\"ur Radioastronomie, Auf dem H\"ugel 69, D-53121 Bonn, Germany\\
$^{3}$SKA Organisation, Jodrell Bank, SK11 9FT, UK\\
}

\date{Accepted XXX. Received YYY; in original form ZZZ}

\pubyear{2020}

\begin{document}
\label{firstpage}
\pagerange{\pageref{firstpage}--\pageref{lastpage}}
\maketitle

\begin{abstract}

The Fast Folding Algorithm (FFA) is a phase-coherent search technique for periodic signals. It has rarely been used in radio pulsar searches, having been historically supplanted by the less computationally expensive Fast Fourier Transform (FFT) with incoherent harmonic summing (IHS). Here we derive from first principles that an FFA search closely approaches the theoretical optimum sensitivity to all periodic signals; it is analytically shown to be significantly more sensitive than the standard FFT+IHS method, regardless of pulse period and duty cycle. A portion of the pulsar phase space has thus been systematically under-explored for decades; pulsar surveys aiming to fully sample the pulsar population should include an FFA search as part of their data analysis. We have developed an FFA software package, \textsc{riptide}, fast enough to process radio observations on a large scale; \textsc{riptide} has already discovered sources undetectable using existing FFT+IHS implementations. Our sensitivity comparison between search techniques also shows that a more realistic radiometer equation is needed, which includes an additional term: the search efficiency. We derive the theoretical efficiencies of both the FFA and the FFT+IHS methods and discuss how excluding this term has consequences for pulsar population synthesis studies.

\end{abstract}



\section{Introduction}

The standard search procedure for periodic pulsar signals relies on taking the Fast Fourier Transform (FFT) of noisy input data and forming the fluctuation power spectrum \citep[see e.g.][for an overview]{PulsarHandbook}. A sufficiently bright periodic signal manifests itself in the power spectrum as a discrete set of \textit{harmonics}: peaks located at integer multiples of the signal's fundamental frequency. The statistical significance of such a signal is evaluated by the sum of its normalized harmonic powers, a practice aptly named \textit{incoherent harmonic summing} \citep[IHS; Huguenin and Taylor, as described by][]{BurnsClark1969}. For each significant candidate thus identified, the original data are then phase-coherently folded modulo the candidate period to produce an integrated pulse profile and other diagnostic information, that are then evaluated by a combination of visual inspection and, in more recent years, machine learning algorithms \citep[e.g.][for a review]{Lyon2016}. The FFT has long dominated the landscape of pulsar searching, being the only periodicity search technique employed in nearly all major pulsar surveys of the past three decades \citep[e.g.][]{Johnston1992, Manchester2001, Cordes2006, Keith2010, Stovall2014}.

However, an alternative method consists of directly folding the data at a wide range of narrowly spaced trial periods; the Fast Folding Algorithm \citep[FFA;][]{Staelin1969} does so efficiently by taking advantage of redundant operations. One integrated profile is produced for each trial period, which can then be evaluated for the presence of a pulse. Although as old as pulsar astronomy itself, for many years the FFA was rarely applied in practice and only for small-scale purposes. Examples include targeted searches of radio emission from X-ray emitting neutron stars \citep{Kondratiev2009, Crawford2009}, or attempts to find an underlying periodicity in the emission of the repeating fast radio burst FRB121102 \citep{Scholz2016}. Outside of pulsar astronomy, we may note that the FFA has been adapted to detect transiting exoplanets in Kepler photometry data \citep{Petigura2013} and used in searches for extra-terrestrial intelligence \citep[SETI@home, e.g.][]{Korpela2001}. When it comes to large-scale FFA searches for pulsars, \citet{Faulkner2004} suggests that one was attempted for some time on the Parkes Multibeam Pulsar Survey \citep{Manchester2001}; they credited it with the discovery of a pulsar with a 7.7s period, but there have been no subsequent publications on the matter. An interest in further developing the FFA has been growing in recent years however, with the practical sensitivity analysis of \citet{Cameron2017} and the implementation of an FFA search pipeline on the PALFA survey by \citet{Parent2018}, that has found a new pulsar missed by the standard FFT pipeline running on the same data.

So far, the FFA has been used mostly in niche cases because of its high computational cost \citep{PulsarHandbook, Cameron2017}. The cost of an FFA search grows unreasonably large if one wants to probe shorter ($P \lesssim 500$ ms) periods, and limited effort has been invested so far in making truly fast FFA implementations. On the other hand, highly optimized FFT libraries are widely available, upon which fast pulsar searching codes can be built. The FFT can also be efficiently extended to search for binary millisecond pulsars \citep[e.g.][]{Ransom2002}. But sensitivity concerns have to be seriously taken into consideration.

Only one theoretical comparison between the sensitivity of the FFA and the standard FFT procedure has been published before \citep{Kondratiev2009}; its conclusion was that the FFA is better at detecting pulsars with both long periods ($P \gtrsim 2$s) and short duty cycles ($\delta \lesssim 1\%$). More recently, a thorough analysis of the sensitivity of the PALFA pulsar survey \citep{Lazarus2015} revealed that FFT searches do not reach their expected theoretical sensitivity to $P \geq 100$ ms pulsars on real-world data, an effect that worsens for longer periods of a few seconds or more. The presence of low frequency ``red" noise was found to be the main cause, and a suggested counter-measure was to use the FFA to search the long period regime. \citet{Cameron2017} focused on the problem of assessing the output of the FFA for the presence of a pulsar signal; they empirically tested the behaviour of several pulse profile evaluation algorithms on a sample of real long-period pulsar observations. Methods based on matched filtering were found to perform better, and the FFA was shown to exceed the sensitivity of the FFT in a number of test cases. \citet{Parent2018} did a similar analysis on artificial data, and reported that on real PALFA survey data, their FFA implementation was on average more sensitive to pulsars with periods longer than a second (see their Fig. 6) than the FFT-based \textsc{presto} pipeline \citep{Ransom2002}. 

Although \textit{empirical} sensitivity comparisons between search methods are important, it is not clear whether the differences in sensitivity thus measured between FFA and FFT search codes demonstrate an intrinsic superiority of either technique in a specific period or duty cycle regime, or if they are indirectly caused by technical implementation details. Particular concerns include how the statistical significance of signals is assessed, or the method used in FFT pipelines to mitigate the effects of low-frequency noise in Fourier spectra, which can have dramatic period-dependent effects \citep{vanHeerden2017}. To fully understand and then realize the true potential of the FFA, the intrinsic sensitivity of search methods must be compared analytically. 

In this paper, we mathematically demonstrate that an FFA search procedure is more sensitive than its FFT counterpart to \textit{all} periodic signals. In \S \ref{sec:algorithm}, we explain the FFA core operation, and propose a more computationally efficient variant of the original FFA algorithm. In \S \ref{sec:sensitivity}, we formulate the general problem of detecting a signal in noisy data as a problem of statistical hypothesis testing; from there, we derive the optimally sensitive test statistic for the presence of a periodic signal in uncorrelated Gaussian noise, and show that calculating it amounts to running the FFA and convolving its output folded profiles with a matched filter. In \S \ref{sec:sensitivity_fft} we use the same hypothesis-testing framework to calculate the sensitivity of the standard FFT+IHS method, and demonstrate that it is always outperformed by the FFA, moreso on smaller pulse duty cycles; pulse period is shown to be irrelevant in this comparison. In \S \ref{sec:implementation}, we describe the details of our own FFA implementation \textsc{riptide}\footnote{\url{https://github.com/v-morello/riptide}} and how we propose to surmount the practical hurdles of realizing the full potential of the FFA on real data contaminated by low-frequency noise and interference. Our implementation is sufficiently fast to be employed on a modern all-sky survey. We discuss the implications of our results in \S \ref{sec:discussion}. In particular, we caution that the radiometer equation of \citet{Dewey1985}, if used to compute the minimum detectable flux density of a search, implicitly assumes that the method employed is phase-coherent and that the pulse shape is known \textit{a priori}; we provide correction factors that more realistically account for the sensitivities of both the FFA and of the incoherent FFT-based standard method. We then conclude in \S \ref{sec:conclusion}.

\section{The Fast Folding Algorithm}
\label{sec:algorithm}

The FFA is an efficient method to phase-coherently fold an evenly sampled time series at multiple closely spaced trial periods at once, taking advantage of the many redundant calculations in the process. It was originally devised to search for periodic signals in one of the very first pulsar surveys \citep{Staelin1969}. 
In this section we explain the principle of the FFA and describe a variant of the algorithm that produces identical results to the original, but does not require any input padding and uses modern CPU caches more efficiently.

\subsection{The folding transform}

\begin{figure}
    \includegraphics[width=\columnwidth]{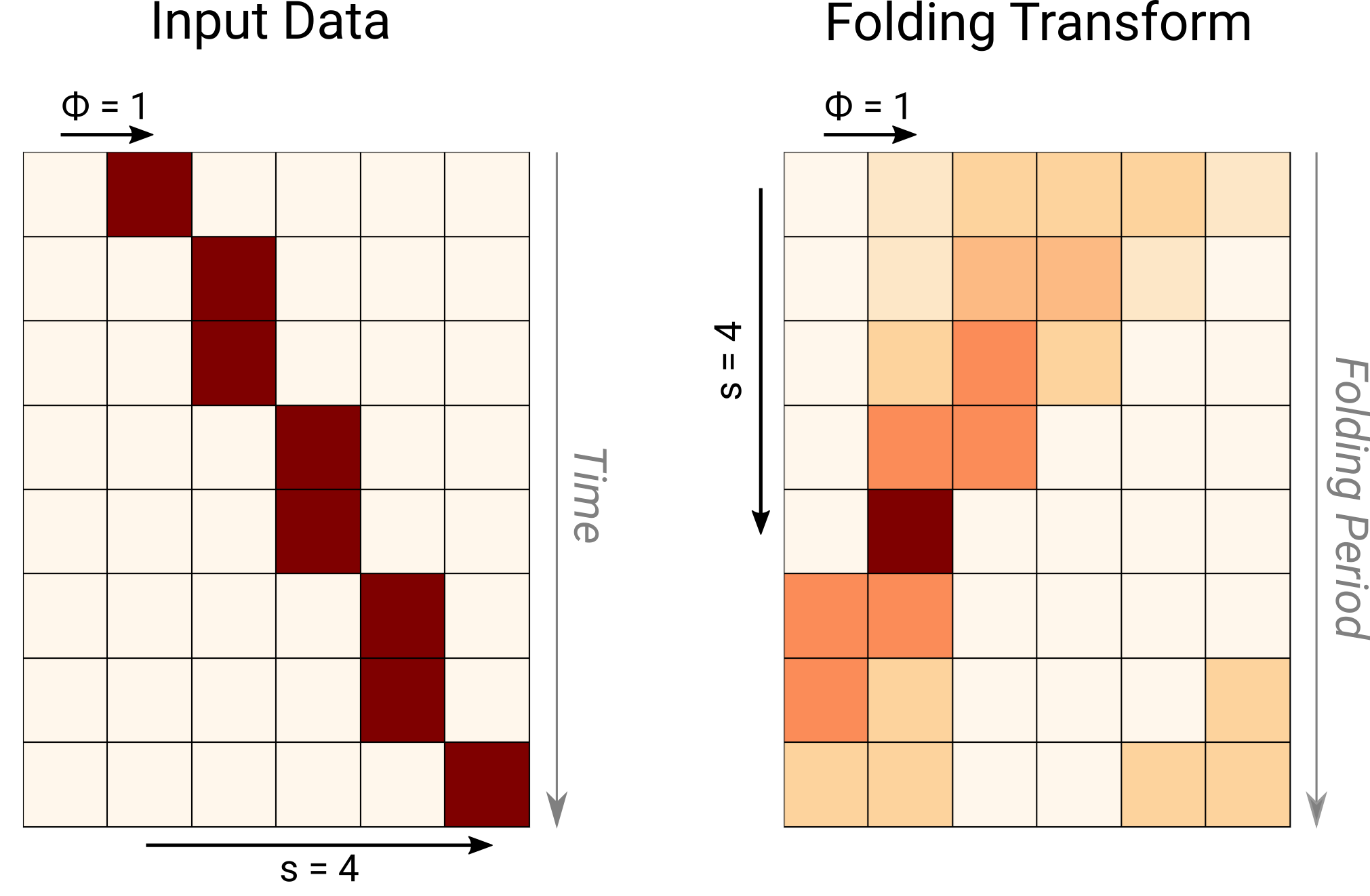}
    \caption{An illustration of the folding transform. Left panel: input data, arranged in $m=8$ rows of $p=6$ samples, containing an artificial train of pulses with a width of one time sample, an initial phase of $\phi=1$ sample and a period larger than $p$, such that the pulses appear to drift in phase by $s=4$ bins over the total observation time. Right panel: folding transform of the input data, where darker colours denote larger integrated intensity. In a blind search, a visible peak in the folding transform denotes a candidate periodic signal; the row and column indices of the peak correspond to its period and initial phase respectively.}
    \label{fig:folding_transform}
\end{figure}

If we consider an evenly sampled time series containing $m$ cycles of a periodic pulsed signal with an integer period $p$ (expressed in number of samples), the data may be represented as a two-dimensional array with $m$ rows and $p$ columns; each row represents an individual cycle, arranged in chronological order from top to bottom, and all pulses appear vertically aligned in phase. In this case, phase-coherently folding the signal is achieved by summing the array across all rows. However, if the period of the signal is increased to a larger, non-integer value $p + \Delta p$ with $0 < \Delta p < 1$, its consecutive pulses will appear to drift in phase towards higher values of pulse phase (i.e. the right-hand side), as illustrated in the example shown in the left panel of Fig. \ref{fig:folding_transform}. To phase-coherently integrate the signal, we now need to apply circular rotations to each pulse in order to compensate for the observed phase drift, which is a linear function of time; in other words, we must integrate the rows of the array along a diagonal path, whose slope can be defined by a single parameter $s$: the number of phase bins drifted towards the right between the first and last rows. Note that the integration path may wrap around in phase at the right edge of the array, possibly multiple times for higher values of $s$. The path $s=0$ corresponds to a folding period of $p$, while $s=m$ corresponds to a folding period of approximately $p+1$. The exact relationship between $s$ and folding period is derived in \S \ref{subsec:ffa_trial_period_spacing}. We call the \textit{folding transform} of the data the set of integrated pulse profiles obtained for all trial values of $s$ with $0 \leq s < m$, which presents itself as an array of the same dimensions as the input data (right panel of Fig. \ref{fig:folding_transform}); column index represents phase, and each row corresponds to a trial folding period between $p$ and approximately $p+1$. Since the input data are discrete in time, integration paths with non-integer values of $s$ are identical to that with the closest integer value, which means that the folding transform provides the highest possible period resolution for the given sampling rate. A brute-force calculation of the folding transform would take $\mc{O}(p m^2)$ additions, but redundancies between consecutive period trials can be exploited.

\subsection{A depth-first traversal, arbitrary length FFA}

\begin{figure*}
    \includegraphics[width=\textwidth]{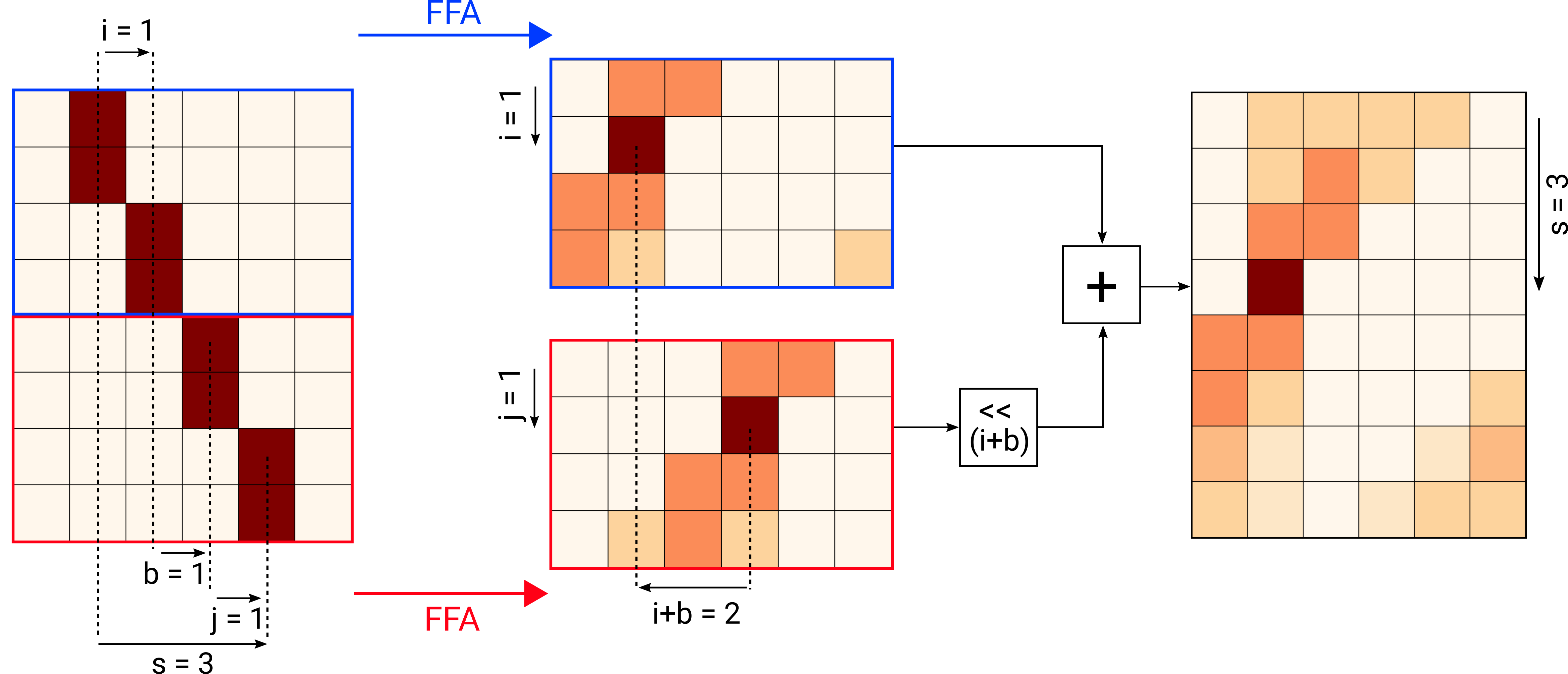}
    \caption{Depth-first variant of the FFA. The principle of the algorithm is to split the input data (left panel) into two parts, outlined in blue and red respectively, and independently calculate their folding transforms (middle panel); by adding together pairs of rows from each of these intermediate outputs with appropriate element-wise circular shifts, the folding transform of the whole input data (right panel) can be obtained. It is shown in this specific example how the fourth row of the output, corresponding to an integration path with a total shift of $s = 3$ bins across the input data, is calculated. See text for a full explanation. This divide and conquer strategy is applied recursively, thus avoiding the redundant additions that would be performed in a brute force calculation of the folding transform.}
    \label{fig:ffa_explained}
\end{figure*}

The FFA can be seen as a divide-and-conquer algorithm, where the idea is to cheaply compute the folding transform of a data block from the folding transforms of its two halves. FFT algorithms are likewise based on the same principle \citep{CooleyTukey65}. Suppose that we have split the input data, in two-dimensional form as in the previous section, into two \textit{arbitrarily} sized blocks that we call the head and the tail, each with $m_h$ and $m_t$ rows respectively such that $m=m_h+m_t$. Let us consider an integration path that shifts to the right by $s$ phase bins between the midpoints in time of the first and last rows: this amounts to $m-1$ rows worth of time span, and the drift rate of the path is thus $s/(m-1)$ phase bins per row. Within the head and tail parts of the data, this integration path will therefore drift by $i$ and $j$ phase bins respectively, whose expressions are
\begin{equation}
\begin{split}
    i &= \nint{(m_h-1)/(m-1) \times s} \\
    j &= \nint{(m_t-1)/(m-1) \times s},
\end{split}
\end{equation}
where $\nint{}$ denotes rounding to the nearest integer. This is illustrated on an example in the left panel of Fig. \ref{fig:ffa_explained}. We further need to take into account a possible 1-bin phase jump at the boundary between the head and the tail that we call $b$:
\begin{equation}
    b = s - (i+j).
\end{equation}
Therefore, if $H$, $T$ and $F$ denote the folding transforms of the head, tail and whole input respectively, then we can obtain the $s$-th row $F_s$ of $F$ by adding together pairs of rows from $H$ and $T$ with the following rule, illustrated on the middle and right panels of Fig. \ref{fig:ffa_explained}:

\begin{equation}
\label{eq:ffa_merge_rule}
    F_s = H_i + \left( T_j \ll (i+b) \right),
\end{equation}
where the notation $\ll k$ denotes the element-wise circular rotation of an array by $k$ elements towards the left; rotating $T_j$ by $i+b$ bins towards the left takes care of phase-connecting the two integration segments. From there a divide-and-conquer algorithm naturally emerges, where the input data are recursively split in two similarly sized blocks, thereby creating a binary tree of folding transforms where the input sizes are (approximately) halved at every recursion level. A leaf node corresponds to an input block with a single row, in which case its folding transform is trivially equal to itself. Our proposed variant of the FFA performs a depth-first, post-order traversal of the tree; the left sub-tree is visited first, resulting in the calculation of $H$, followed by the traversal of the right sub-tree, resulting in the calculation of $T$, and then the two are combined into $F$ using Eq. \ref{eq:ffa_merge_rule}. The original variant of the FFA described by \citet{Staelin1969} instead performs a breadth-first traversal from the bottom-level of the tree: first, the input is partitioned in two-row blocks and their folding transforms are calculated; from there, the folding transforms of all four-row blocks are built from the results of the previous step, and so on until the transform of the whole input data has been obtained.



The depth-first traversal FFA presented here is more efficient than the original breadth-first variant for two reasons. Firstly, the depth-first variant can directly handle inputs with an arbitrary, non power-of-two number of rows $m$, removing the need to zero-pad the input data. This not only saves operations in the folding transform calculation itself, but also avoids creating additional folded profiles to be evaluated for the presence of a pulse. We note that the input data must still be presented in rectangular form, with a total number of samples that is a multiple of $p$. If the last input row is incomplete, it must be either ignored or padded to a full $p$ samples; we have chosen the former option in our implementation because it is more computationally efficient. Secondly, a depth-first traversal of the tree has better memory locality and is more optimally suited to the hierarchical cache structure of modern CPUs: as the traversal proceeds towards the deeper layers of the tree, the calculations associated with a node and all of its sub-tree can fit entirely in cache once a certain depth is reached, regardless of the size of the full input. In contrast, the original breadth-first variant has to initially partition the input into two-row blocks and compute their respective transforms; for large inputs that cannot fit in the CPU cache, this requires two comparatively slow transactions with the main memory (RAM): one to read the entire input, another to write an equal-sized intermediate output with all two-row blocks transformed. All these operations have to be done \textit{before} computing the next intermediate output with four-row blocks transformed. Building the final result in this order always requires a total of $2 \log_2(m)$ RAM transactions, which negatively affects execution times.

\subsection{Spacing of trial periods}
\label{subsec:ffa_trial_period_spacing}


The statement in \citet{Staelin1969} that the trial periods of the FFA are spaced linearly is true only as a first-order approximation.
Let $\tau$ denote the sampling time, $P_0 = p \tau$ the chosen base folding period and $\nu_0 = 1/P_0$ the corresponding base folding frequency. Between the midpoints in time of the first and last rows of the input array, the total time elapsed is $T = (m-1) p \tau$. If the data contain a signal with a different frequency $\nu_0 + \Delta \nu$, then between the first and last rows, its train of pulses will drift in phase by some number of bins $s$ to the right. The relationship between $s$ and $\Delta \nu$ is obtained by writing the phase difference of the pulse train (with frequency $\nu$) relative to the folding frequency $\nu_0$. This phase difference, expressed in number of folding periods, is given by $\Delta \phi(t) = (\nu - \nu_0) t$. At $t = T$, the pulse train is \textit{late} in phase by $s$ time samples with respect to the fiducial point which means that $\Delta \phi(T) = -s/p$ and thus

\begin{equation}
\label{eq:frequency_trial_spacing}
\Delta \nu = -\frac{s}{p} \times \frac{1}{T}.
\end{equation}
The output profiles produced by the FFA are therefore spaced linearly \textit{in frequency} rather than in period. The $s$-th row in the folding transform therefore corresponds to a trial frequency of

\begin{equation}
\label{eq:kth_frequency_trial}
\nu_s = \left( \frac{1}{p} - \frac{s}{(m-1) p^2} \right) \times \frac{1}{\tau} ,
\end{equation}
and to a trial period of

\begin{equation}
\label{eq:kth_period_trial}
P_s = \frac{p^2}{p - s/(m-1)} \times \tau.
\end{equation}
It is worth noting that the last period trial $P_{m-1}$ is strictly larger than $(p+1) \tau$. In a practical search, where a sequence of folding transforms with increasing base periods are calculated, this means that their last few rows may have trial periods larger than the base period of the next folding transform. These redundant trials are ignored in our search code implementation.

\section{Optimally searching for periodic signals and sensitivity of the FFA}
\label{sec:sensitivity}

In this section we examine the general problem of testing for the presence of a signal in discretely sampled noisy data, which is a matter of statistical hypothesis testing. After introducing all necessary technical terms, we find from first principles the optimally sensitive test statistic for detecting a signal with known parameters. For periodic signals, it emerges that computing this test statistic is equivalent to phase-coherently folding the data and correlating the output with a matched filter that reproduces the known pulse shape. In a practical search the signal parameters are \textit{not} known \textit{a priori}; in this case a sensitivity penalty is incurred, which we quantify.

\subsection{Problem formulation}
\label{subsec:hypothesis_testing_formulation}




In ideal observing conditions, the evenly sampled time series data to search can be represented by an $N$-dimensional vector $\mb{x}$ resulting from the sum of background noise $\mb{w}$ and possibly an astrophysical signal $\mb{s}$, which can be written as

\begin{equation}
\label{eq:input_data}
\mb{x} = a \mb{s} + \mb{w}.
\end{equation}
The samples of $\mb{w}$ are assumed to be mutually independent and drawn from the same normal distribution with \textit{known} mean $m$ and standard deviation $\sigma$; we will consider the case $m = 0$ and $\sigma = 1$ without loss of generality, as the data can be normalised beforehand with the transformation $\mb{x} \mapsto (\mb{x} - m) / \sigma$. $a \geq 0$ is an unknown parameter that denotes the amplitude of the signal, noting that we impose a \textit{unit square sum} (or unit L2-norm) normalization on $\mb{s}$:

\begin{equation}
\label{eq:signal_normalisation}
\begin{split}
\sum_{i=1}^{N} s_i^2 &= \mb{s} \cdot \mb{s} = 1,
\end{split}
\end{equation}
where the dot symbol denotes the dot product. Why the choice of unit square sum normalisation is natural here will become apparent later. We will assume, until stated otherwise, that the shape of the astrophysical signal is known in advance, i.e. we are searching for a well-defined signal shape (the value of all $s_i$ are known) with unknown amplitude. Establishing whether or not the signal $\mb{s}$ is present in $\mb{x}$ means deciding between the mutually exclusive statistical hypotheses below:

\begin{itemize}
\item $H_0$: the data $\mb{x}$ contain no signal, $a = 0$.
\item $H_1$: the data $\mb{x}$ contain the signal $\mb{s}$, $a > 0$.
\end{itemize}
To that end, we need a mathematical function $T(\mb{x})$, a \textit{test statistic}, that reduces the data $\mb{x}$ to a single summary value from which one can sensibly decide between hypotheses. For any test statistic, one must also define a so-called \textit{critical region}, which is the set of values of $T(\mb{x})$ for which the null hypothesis is rejected. In most practical applications, as will be the case here, the critical region will simply be defined by a threshold above which (or below which) one decides in favour of $H_1$. Here we want to find a test statistic (and critical threshold) that maximizes the sensitivity of our search for the presence of $\mb{s}$, a concept that first needs to be rigorously defined.

\subsection{Evaluating a test statistic}

%
%

When choosing between hypotheses, two types of errors can be made. A type I error is a rejection of the null hypothesis $H_0$ when it is true, which in pulsar astronomy parlance is a false alarm or false positive. A type II error is a rejection of the alternative hypothesis $H_1$ when it is true $-$ here, quite simply a missed pulsar. Unless the decision problem is trivial, there is no perfectly accurate test, so that there exists a trade-off between the two types of error rates; when evaluating tests, the standard practice is therefore to first set an acceptable false positive rate, or \textit{significance level} noted $\alpha$, and identify which test exhibits the lowest type II error rate while maintaining a type I error probability smaller than $\alpha$ \citep[e.g.][]{StatisticalInference}. For our specific purposes, the sensitivity of a test statistic $T$ to a given signal shape $\mb{s}$ can be rigorously evaluated by following the procedure below:

\begin{enumerate}
\item Set the significance level $\alpha$. In a pulsar search, the choice of $\alpha$ is usually made so that no more than one false alarm per observation occurs on average, assuming ideal data without interference signals \citep[e.g.][]{PulsarHandbook}.

\item Calculate the critical threshold $\eta$ of the test statistic $T$ associated with the chosen false alarm rate $\alpha$. $\eta$ is defined by
\begin{equation}
\label{eq:significance}
P(T(\mb{x}) \geq \eta | H_0) = \alpha,
\end{equation}
where the notation $P(E|H)$ denotes the probability of event $E$ given hypothesis $H$ is true.
\item Assuming the alternative hypothesis $H_1$ is true, compute the probability $\gamma$ that $T$ exceeds the critical value:
\begin{equation}
\label{eq:statistical_power}
P(T(\mb{x}) \geq \eta | H_1) = \gamma(a).
\end{equation}
$\gamma$ is called the \textit{statistical power} of the test $T$, or \textit{completeness} of the search, and it is a function of the signal amplitude $a$.
\end{enumerate}
Note that we have implicitly assumed that $T$ is a monotonically increasing function of $a$. If instead $T$ is a monotonically decreasing function of $a$, the inequalities in \ref{eq:significance} and \ref{eq:statistical_power} must be reversed.
One last useful concept is the \textit{upper limit} $\mc{U}$, which is the minimum signal amplitude that is detectable with at least some probability $\gamma_{\rm{min}}$, following the rigorous definition of \citet{Kashyap2010}. It is a characteristic of any search procedure and is defined by
\begin{equation}
\label{eq:upper_limit}
\gamma(\mc{U}) = \gamma_{\rm{min}}.
\end{equation}
For the sake of simplicity we will set $\gamma_{\rm{min}} = 0.5$ throughout the paper. This now provides us with a rigorous definition of sensitivity: we can conveniently compare two test statistics by their respective upper limits at some fixed significance level $\alpha$.

\subsection{Constructing the most powerful test: the Z-statistic}
\label{subsec:z_statistic}


One interesting and widely applied test is the so-called likelihood-ratio test (LRT) statistic $\Lambda$, whose most general definition is \citep[e.g.][]{StatisticalInference}:
\begin{equation}
\label{eq:likelihood_ratio_test_definition}
\Lambda(\mb{x}) = \frac{ \sup_{\theta \in \Theta_0} \mc{L}(\theta |\mb{x}) }{ \sup_{\theta \in \Theta} \mc{L}(\theta |\mb{x}) },
\end{equation}
where $\theta$ is a vector of model parameters, $\mc{L}(\theta |\mb{x})$ the likelihood function, $\Theta$ is the entire parameter space, $\Theta_0$ the subset of that space spanned by the null hypothesis $H_0$ and $\rm{sup}$ denotes the supremum. The LRT takes values between zero and one and essentially measures the goodness-of-fit of the null hypothesis. The main reason to investigate the LRT is that under some specific conditions it has optimal statistical power; furthermore, writing out its expression (if possible), may also suggest another equally powerful test with a simpler expression. For our particular problem, we have a single parameter $a$ which only takes positive values, and the natural logarithm of $\Lambda(\mb{x})$ reduces to a simple form
\begin{equation}
\label{eq:likelihood_ratio_test}
\ln \Lambda(\mb{x}) = -\frac{1}{2}(\mb{x} \cdot \mb{s})^2,
\end{equation}
which is derived in the appendix (\S \ref{subsec:lrt_expression}). The above expression depends on the data only through what we will call the Z-statistic
\begin{equation}
\label{eq:z_statistic}
Z(\mb{x}) = \mb{x} \cdot \mb{s} = \sum_{i=1}^{N} s_i x_i,
\end{equation}
with the intuition that it will retain all desirable properties of the LRT. The Z-statistic is in fact the maximum-likelihood estimator for the parameter $a$ (see \S \ref{subsec:lrt_expression}) and it follows a normal distribution with mean $a$ and unit variance, which we can remark by expanding the above:
\begin{equation}
\label{eq:z_statistic_expansion}
\begin{split}
Z(\mb{x}) &= \mb{x} \cdot \mb{s} \\
          &= a \mb{s} \cdot \mb{s} + \mb{w} \cdot \mb{s} \\
          &= a + \sum_{i=1}^{N} s_i w_i.
\end{split}
\end{equation}
By our initial assumption, the $w_i$ are independent variables that follow a \textit{standard} normal distribution, that is with zero mean and unit variance, which is compactly written as $w_i \sim \mc{N}(0,1)$. In the above equation, the sum of the $w_i$ weighted by the coefficients $s_i$ is also a normal random variable, whose mean is zero (by linearity of expectation) and whose variance is equal to the sum of the $s_i^2$, which is 1 by definition (Eq. \ref{eq:signal_normalisation}). Therefore $Z(\mb{x}) \sim \mc{N}(a, 1)$.

From there we may find the critical value of the Z-statistic for a given significance level $\alpha$ by applying Eq. \ref{eq:significance}. Under the null hypothesis, we have $a = 0$ and thus we can write $P(Z(\mb{x}) \geq u | H_0) = \overline{\Phi}(u)$ where $\overline{\Phi}$ is the survival function of the standard normal distribution $\mc{N}(0,1)$. The critical value for the Z-statistic $\eta_z$ at significance level $\alpha$ is the solution to the equation $\overline{\Phi}(u) = \alpha$, and its expression thus given by 

\begin{equation}
\label{eq:z_statistic_critval}
\eta_z(\alpha) = \overline{\Phi}^{-1}(\alpha),
\end{equation}
where $\overline{\Phi}^{-1}$ denotes the inverse survival function of the standard normal distribution. $\eta_z$ therefore represents the number of Gaussian sigmas associated with the probability $\alpha$. By application of the Karlin-Rubin theorem, the test defined as follows:
\begin{itemize}
    \item Choose $H_1$ if $Z(\mb{x}) \geq \eta_z(\alpha)$
    \item Choose $H_0$ otherwise
\end{itemize}
is the most powerful test for the presence of the signal $\mb{s}$ regardless of the chosen significance level $\alpha$. A demonstration is provided in the appendix (\S \ref{subsec:z_statistic_optimality_demo}). Its statistical power as a function of true signal amplitude $a$ can be found directly from Eq. \ref{eq:statistical_power} remarking that, under the alternative hypothesis $H_1$, $Z(\mb{x}) - a$ follows a standard normal distribution:
\begin{equation}
\gamma_z(a) = \overline{\Phi}(\eta_z - a) = \Phi(a - \eta_z),
\end{equation}
where $\Phi$ denotes the cumulative distribution function of the standard normal distribution and where the property $\overline{\Phi}(u) = \Phi(-u)$ has been used. We can remark that for $a = \eta_z$, $\gamma_z = \Phi(0) = 0.5$ and therefore from Eq. \ref{eq:upper_limit} that the upper limit of the Z-test is simply
\begin{equation}
\label{eq:z_statistic_upperlim}
\mc{U}_z = \eta_z = \overline{\Phi}^{-1}(\alpha),
\end{equation}
which is thus the lowest achievable at significance level $\alpha$ among all statistical tests. Since the Z-statistic is the best estimator of the signal amplitude relative to the background noise standard deviation (here set to 1 without loss of generality), we will interchangeably call it \textit{signal-to-noise ratio} (SNR or S/N) when later discussing its practical software implementation.

\subsection{Case of periodic signals}
\label{subsec:case_of_periodic_signals}

If we further assume that the signal $\mb{s}$ is periodic with a known integer period of $p$ samples, then $\mb{s}$ is a sequence of identical segments $\mb{t}$ that we will call the pulse template. In this case, it is equivalent to calculate the Z-statistic by phase-coherently folding the data $\mb{x}$ at a period $p$ into a so-called folded profile $\mb{f}$, and then by taking the dot product of the folded output with the pulse template $\mb{t}$. Assuming that the data contain exactly $m$ pulse periods, the elements of $\mb{f}$ are given by

\begin{equation}
    f_j = \sum_{i=0}^{m-1} x_{pi+j},
\end{equation}
where $0 \leq i < m$ represents the pulse index and $0 \leq j < p$ the phase bin index. Since $\mb{s}$ is periodic the elements of the pulse template are such that $t_j = s_{pi+j}$ for any $i, j$. Writing out the Z-statistic expression and rearranging, we get

\begin{equation}
\label{eq:z_statistic_folded}
\begin{split}
\mb{x} \cdot \mb{s} &= \sum_{i=0}^{m-1} \sum_{j=0}^{p-1} x_{pi+j} s_{pi+j} \\
    &= \sum_{i=0}^{m-1} \sum_{j=0}^{p-1} x_{pi+j} t_j \\
    &= \sum_{j=0}^{p-1} t_j \sum_{i=0}^{m-1} x_{pi+j} \\
    &= \sum_{j=0}^{p-1} f_j t_j \\
    &= \mb{f} \cdot \mb{t}
\end{split}
\end{equation}
which completes our demonstration. When searching for a periodic signal with a known period and pulse shape, the optimally sensitive procedure to detect its presence is based upon phase-coherently folding the input data into a profile and then correlating it with the known pulse template.

\subsection{Searching for periodic signals with unknown parameters}

Until now, we have considered the \textit{true} signal $\mb{s}$ to be known in advance; this is not the case in a blind search, where one must make \textit{a priori} assumptions about the properties of the signal. We may end up testing the input data, erroneously so, for the presence of a different trial signal $\mb{s}'$, which means using a less powerful test statistic $Z'$:
\begin{equation}
    Z'(\mb{x}) = \mb{x} \cdot \mb{s'} = a \mb{s} \cdot \mb{s'} + \mb{w} \cdot \mb{s'},
\end{equation}
where $\mb{s'}$ is also normalised to unit square sum. The relevant term in the equation above is what we will call the \textit{search efficiency}
\begin{equation}
\label{eq:efficiency}
    \mc{E} = \mb{s} \cdot \mb{s'},
\end{equation}
which measures the similarity between the true and postulated signals, noting that $|\mc{E}| \leq 1$ and that $\mc{E} = 1$ if and only if $\mb{s'} = \mb{s}$. Following the same steps as in \S \ref{subsec:z_statistic}, one can show that the degraded test statistic Z' follows a normal distribution $\mc{N}(\mc{E} a, 1)$; its critical value remains equal to $\eta_z$, but its statistical power is reduced to a value of
\begin{equation}
\gamma_z'(a) = \Phi(\mc{E} a - \eta_z),
\end{equation}
and its correspondingly higher upper limit $\mc{U}_z'$ is given by
\begin{equation}
\label{eq:upper_limit_degraded}
\mc{U}_z' = \frac{1}{\mc{E}} \mc{U}_z.
\end{equation}
One therefore needs to search for a parametrized family of signal templates so that at least one is acceptably similar to the true signal. Periodic pulsar signals need to be described by at least three parameters:
\begin{itemize}
    \item Their period $p$
    \item Their pulse phase $\phi$ at the start of the observation
    \item Their duty cycle $\delta$, defined as the ratio between pulse width and period
\end{itemize}
This naturally raises the question of gridding this three-dimensional search space so that the efficiency $\mc{E}$ remains acceptably high. We also need to estimate $\mc{E}$ in a practical FFA search. The FFA conveniently provides the finest possible period coverage allowed by the time resolution of the data, by producing a sequence of folded profiles at all distinguishable trial periods. Using the result of \S \ref{subsec:case_of_periodic_signals}, these profiles can be directly tested for the presence of various pulse templates, parameterized by a duty cycle $\delta$ and initial phase $\phi$. However, the finest possible coverage of the phase dimension can be achieved using matched filtering; that is performing circular convolution of the profiles with templates with different values of $\delta$ and an initial phase of zero. This only leaves the gridding of the duty cycle dimension for detailed consideration; since $\delta$ can span several orders of magnitude in known pulsars, an efficient strategy is to geometrically space the trial duty cycle values, where the minimum trial value corresponds to a width of one phase bin. To provide an estimate of $\mc{E}$ in a real search using this strategy, we need to postulate a shape both for the true signal and the templates (or matched filters). This investigation can be highly multi-dimensional; we will assume that the true pulse shape is Gaussian, which is a good approximation for the majority of pulsars \citep[e.g.][]{Rankin1983}, and that the templates are so-called boxcar or top-hat filters because they are very computationally efficient in practice (\S \ref{subsec:matched_filtering}). Under these assumptions, and if the trial boxcar widths are geometrically spaced by a factor of 2, the average efficiency when searching for a Gaussian pulse with \textit{unknown} duty cycle is 

\begin{equation}
    \mc{E}_\mathrm{FFA} \approx 0.930,
\end{equation}
which we derive in the appendix (\S \ref{subsec:boxcar_efficiency}). This can be taken as a reasonable approximation of the average efficiency of a practical FFA search, and shows that for single-peaked profiles boxcars provide quite acceptable sensitivity. For more exotic, multi-peaked pulsar profiles however, the above is an overestimate. We leave the problem of choosing the optimal set of matched filters with respect to the known pulsar population open for future investigation. From Eq. \ref{eq:upper_limit_degraded}, the upper limit of a practical FFA search using boxcar filters at a significance level $\alpha$ can be approximated by

\begin{equation}
\label{eq:ffa_search_practical_upperlim}
    \mc{U}_\mathrm{FFA} \approx 1.075~ \overline{\Phi}^{-1}(\alpha),
\end{equation}
or in other words, 1.075 times the number of Gaussian sigmas associated with the probability $\alpha$.

\section{Sensitivity analysis of the standard FFT procedure}
\label{sec:sensitivity_fft}

After establishing the optimal, or near-optimal sensitivity of an FFA search procedure, we now need to quantify how the standard incoherent FFT-based search performs in comparison. Although the topic of the sensitivity of a sum of Fourier harmonics to periodic signals has been covered before \citep[e.g.][]{Groth1975, Vaughan1994, Ransom2002, Kondratiev2009}, we will do so again in a way that fits the hypothesis testing framework laid out in the previous section. This will allow a mathematically rigorous comparison of FFA and FFT upper limits.

\subsection{The incoherent detection statistic}

For the purposes of this section, we will use the so-called unitary normalisation of the discrete Fourier transform (DFT); the DFT of a sequence $\mb{x}$ with $N$ elements is the set of coefficients $\mb{X}$ whose elements are given by
\begin{equation}
\label{eq:unitary_dft}
    X_k = \frac{1}{\sqrt{N}} \sum_{n=0}^{N-1} x_n \exp \left( -2 i \pi k \frac{n}{N} \right),
\end{equation}
where $0 \leq k < N$ and $i^2=-1$. We will use uppercase letters to denote the complex-valued, unitary Fourier transforms of the vectors defined in Eq. \ref{eq:input_data}. The unitary DFT conserves the complex dot product and therefore the L2-norm, such that in particular $\norm{\mb{S}}^2 = \norm{\mb{s}}^2 = 1$. Because the $w_i$ are mutually independent and follow a standard normal distribution, the real and imaginary parts of all complex elements of $\mb{W}$ are also mutually independent and normally distributed with zero mean and variance 1/2; this important property will be relevant below \citep[a demonstration of it can be found in e.g. Chapter 15 of][]{Kay93}. We will assume that $\mb{s}$ is a periodic signal whose fundamental frequency is $k_0/N$ where $k_0$ is a positive integer. In this case, all elements of $\mb{S}$ are zero, except for indices that are integer multiples of $k_0$ which contain all of the signal's power \citep[e.g.][]{Ransom2002}. The incoherent detection statistic $I$ is defined as the sum of harmonic powers up to some rank $h$:
\begin{equation}
\label{eq:ids_definition}
\begin{split}
    I(a, h) &= \sum_{i=-h}^{h} |X_{k_0 i}|^2 \\
            &= |X_0|^2 + 2 \sum_{i=1}^{h} |X_{k_0 i}|^2,
\end{split}
\end{equation}
where we have used the Hermitian symmetry of the DFT of a real-valued input. In a practical search, the optimal number of harmonics to sum $h$ is unknown \textit{a priori} and depends mainly on the duty cycle of the pulse. This requires trying multiple values of $h$, and therefore calculating multiple incoherent detection statistics so that sensitivity is maximized to a broader range of pulse shapes. In practice, trial values of $h$ are powers of two up to either 16 or 32.

\subsection{Distribution of the incoherent detection statistic}

Let us first define, for practical purposes, the fraction of power contained in the Fourier harmonics of the postulated signal $\mb{s}$ up to rank $h$:
\begin{equation}
\begin{split}
    P(h) &= \sum_{i=-h}^{h} |S_{k_0 i}|^2 \\
         &= |S_0|^2 + 2 \sum_{i=1}^{h} |S_{k_0 i}|^2,
\end{split}
\end{equation}
which is equal to 1 when all harmonics of the signal are summed. For each element of $\mb{X}$, with the exception of $X_0$, we can write

\begin{equation}
\label{eq:xi_squared_expansion}
\begin{split}
2 |X_i|^2 &= 2 |a S_i + W_i|^2  \\
          &= |a \sqrt{2} S_i + \sqrt{2} W_i|^2  \\
          &= (a \sqrt{2} \Re{S_i} + \sqrt{2} \Re{W_i})^2 \\
          &~+ (a \sqrt{2} \Im{S_i} + \sqrt{2} \Im{W_i})^2.
\end{split}
\end{equation}
Here we must recall that $\Re{W_i}$ and $\Im{W_i}$ are independent random variables that follow a normal distribution with mean 0 and variance 1/2, while $S_i$ is a complex-valued constant. This means that the terms $\sqrt{2} \Re{W_i}$ and $\sqrt{2} \Im{W_i}$ follow $\mc{N}(0,1)$. The equation above is thus the sum of squares of two independent normal variables with unit variances, and means equal to $a \sqrt{2} \Re{S_i}$ and $a \sqrt{2} \Im{S_i}$ respectively. $2 |X_i|^2$ therefore follows, by definition, a \textit{noncentral chi-squared distribution}:

\begin{equation}
    2 |X_i|^2 \sim \chi^2 \left(d=2, \lambda=2 a^2 |S_i|^2) \right)
\end{equation}
where $d$ denotes degrees of freedom and $\lambda$ is the so-called noncentrality parameter: it is defined as the sum of the squared means of every contributing normal variable. The case of $X_0$ must be treated separately; writing out its expression from Eq. \ref{eq:unitary_dft} shows that it is real-valued and normally distributed with variance 1 instead of 1/2. Thus $|X_0|^2 \sim \chi^2 \left(d=1, \lambda=a^2 |S_0|^2 \right)$. From there, we can simply expand each term of Eq. \ref{eq:ids_definition} into its real and imaginary part as we did in Eq. \ref{eq:xi_squared_expansion}, remark that all random variables involved are mutually independent, and conclude that $I(a, h)$ likewise follows the noncentral chi-squared distribution below:\footnote{\citet{Groth1975} provides, with a different formalism, series expansions for the PDF and CDF of this distribution but does not mention its name.}
\begin{equation}
\label{eq:ids_distribution}
    I(a, h) \sim \chi^2 \left(d=2h+1, \lambda=a^2 P(h) \right).
\end{equation}
The distribution of the incoherent detection statistic is fully specified by three parameters: the number of harmonics summed, the true amplitude of the signal, and its cumulative harmonic power distribution $P(h)$ which can be numerically calculated by taking the DFT of any postulated pulse shape. It is important to note that the distribution does \textit{not} depend on the signal period, and so will be the case of any relative differences in sensitivity between FFA and FFT.

\subsection{Statistical power analysis}

\begin{figure*}
\centering
\includegraphics[width=1.00\textwidth]{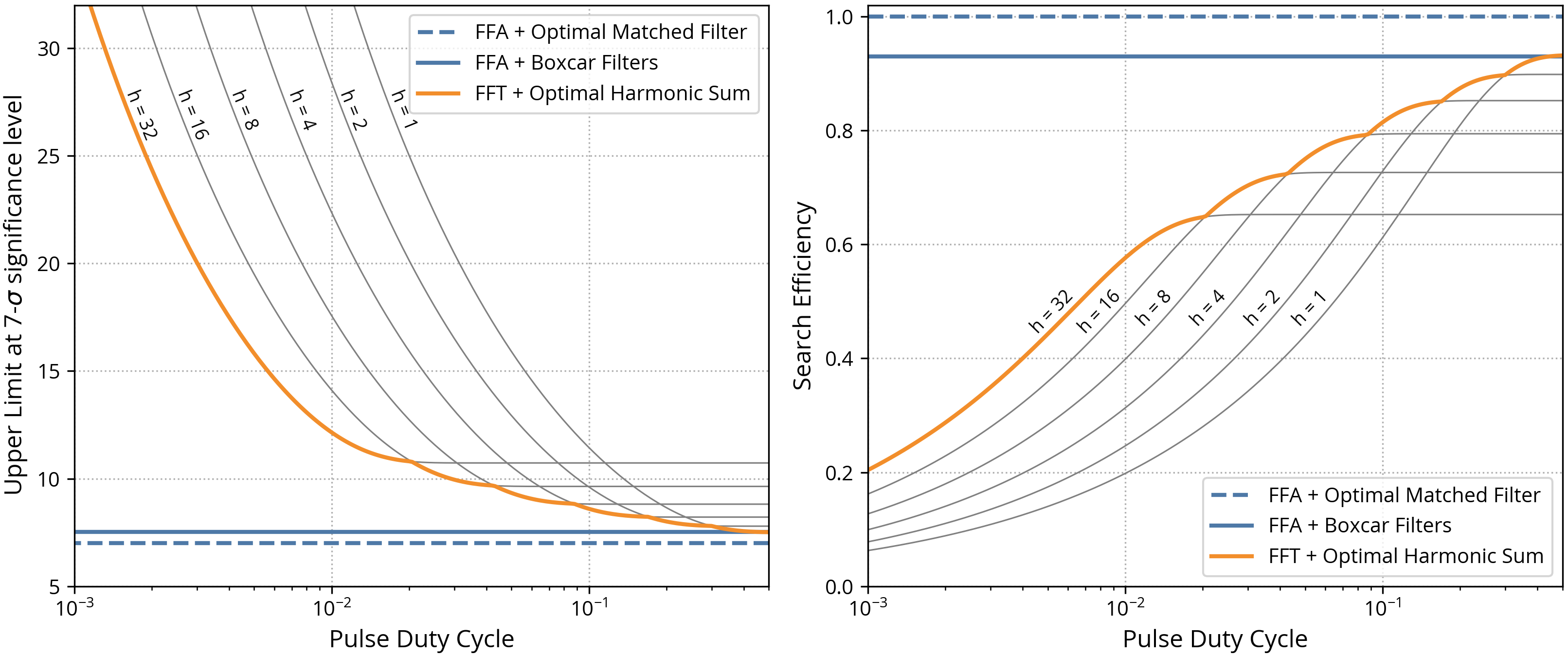}
\caption{Sensitivity comparison between an FFA-based search and the standard FFT procedure with incoherent harmonic summing. We assumed a significance threshold of 7-$\sigma$ and a periodic signal with a Gaussian pulse shape. Left panel: upper limit as a function of Gaussian pulse FWHM (in log-scale), expressed in units of pulse period. The upper limit of a search procedure is the minimum signal amplitude (in units of background noise standard deviation) that it can detect with 50\% probability. See \S \ref{subsec:hypothesis_testing_formulation} for a rigorous definition of amplitude. The theoretical optimum upper limit ($\mc{U}_z = 7$ here) could be achieved by the FFA if the pulse shape and width were known \textit{a priori} (blue dashed line); in practice, a penalty is incurred when using boxcar matched filters (blue solid line). FFT search codes compute multiple harmonic sums, typically up to 32 harmonics, whose individual upper limits are plotted as grey curves; the upper limit of the whole FFT procedure corresponds to their lower envelope (orange line). Right panel: search efficiency $\mc{E}$, i.e. ratio between the theoretical optimum upper limit $\mc{U}_z$ and the upper limit of each search procedure. $\mc{E}_\mathrm{FFA} \approx 0.93$ in practice. It is implicitly assumed that the FFA search employs enough output phase bins to resolve the pulse.}
\label{fig:sensitivity_comparison}
\end{figure*}

With the distribution of the detection statistic $I$ now known, we only need to follow the procedure laid out in the previous section to determine its sensitivity to any periodic signal. A number of software packages provide routines accurate to machine precision that compute the PDF, CDF and survival function (as well as their respective inverse functions) of the noncentral chi-squared distribution; we used the \textsc{python} language implementations of the \textsc{scipy.stats} package \citep{scipy} in this section. For a given number of harmonics summed $h$, the critical value $\eta_I(h)$ of the incoherent detection statistic is obtained from Eq. \ref{eq:significance}, via the application of the inverse survival function of the chi-squared distribution with $d = 2h+1$ degrees of freedom $\overline{F}^{-1}_{\chi^{2}(2h+1)}$:

\begin{equation}
    \eta_I(h) = \overline{F}^{-1}_{\chi^{2}(2h+1)}(\alpha).
\end{equation}
The completeness $\gamma_I$ (at significance level $\alpha$) is a function of both signal amplitude $a$ and number of harmonics summed $h$, and is likewise found using Eq. \ref{eq:statistical_power}:

\begin{equation}
\label{eq:ids_statistical_power}
    \gamma_I(a, h) = \overline{F}_{\chi^{2}(2h+1, \lambda(a))} \left( \eta_I(h) \right),
\end{equation}
where $\overline{F}_{\chi^{2}(2h+1, \lambda(a))}$ is the survival function of the noncentral chi-squared distribution of Eq. \ref{eq:ids_distribution}. The upper limit of the incoherent detection statistic is the true signal amplitude $\mc{U}_I(h)$ such that the signal is detected with probability 0.5 in a sum of $h$ harmonics, as defined in Eq. \ref{eq:upper_limit}:

\begin{equation}
    \gamma_I \left( \mc{U}_I(h), h \right) = 0.5.
\end{equation}
Calculating $\mc{U}_I(h)$ means solving for $a$ the equation $\gamma_I(a, h) = 0.5$; $\gamma_I$ is a monotonically increasing function of $a$ which means that the equation can be numerically solved quickly using the bisection method, an implementation of which is available in the \textsc{scipy.optimize} package. Obtaining the value of $\mc{U}_I(h)$ as a function of pulse duty cycle and different values of $h$ now allows us to directly compare the sensitivity of the standard FFT search method with that of the FFA, for the same fixed significance level $\alpha$. We need to set two free parameters to complete this comparison:
\begin{itemize}
    \item The pulse shape, which defines the cumulative harmonic power distribution $P(h)$: to remain consistent with the analysis done on the FFA above we also use Gaussians, parametrized by their full-width at half-maximum (FWHM) in units of pulse period
    \item The significance level, that we set to 7-$\sigma$, i.e. $\alpha = \bar{\Phi}(7)$ where $\bar{\Phi}$ is the survival function of the normal distribution.
\end{itemize}
We have plotted the upper limits $\mc{U}_I(h)$ for $h = 1, 2, ..., 32$ in the left panel of Fig. \ref{fig:sensitivity_comparison}, as a function of Gaussian pulse FWHM: these different upper limits are shown as a set of annotated grey lines. For each pulse duty cycle, there is one optimal harmonic sum, which is that with the smallest upper limit and therefore highest sensitivity. The upper limit of the standard FFT search procedure, assuming that up to 32 harmonics are summed, is therefore

\begin{equation}
    \mc{U}_\mathrm{FFT} = \min_{h = 1, 2, ..., 32} \mc{U}_I(h),
\end{equation}
which is represented by the orange line on the left-hand side of Fig. \ref{fig:sensitivity_comparison}. The ratios $\mc{U}_\mathrm{z} / \mc{U}_\mathrm{I}(h)$ and $\mc{U}_\mathrm{z} / \mc{U}_\mathrm{FFT}$ are directly comparable to the \textit{search efficiency} defined in the previous section, which is evident from Eq. \ref{eq:upper_limit_degraded}; these ratios are plotted on the right panel of Fig. \ref{fig:sensitivity_comparison} and provide a more convenient comparison of search methods. The optimal upper limit $\mc{U}_z$ can only be achieved by the FFA procedure if the pulse shape and width are known in advance, and thus we have also plotted the more realistic upper limit for a practical FFA search $\mc{U}_\mathrm{FFA}$, given by Eq. \ref{eq:ffa_search_practical_upperlim}.

We note that some past surveys \citep[e.g.][]{Manchester2001} have used the phase information of the harmonics of candidates reported by the incoherent harmonic sums to reconstruct their integrated pulse profile, and also estimate their phase-coherent S/N. Although this process is useful in removing chance detections, it does \textit{not} change the upper limit of the search as a whole for a fixed false alarm probability $\alpha$, because it can only be applied to signals bright enough to exceed the detection threshold of the FFT+IHS procedure in the first place. Computing resources permitting, implementing this second-stage selection process might however justify an increase of the acceptable $\alpha$ for the FFT+IHS procedure in practice.

\subsection{Summary}

We have rigorously shown that in the presence of pure uncorrelated Gaussian noise, a phase-coherent FFA-based search is more sensitive than the incoherent FFT search procedure regardless of pulse period and duty cycle. The improvement offered by the FFA is greater for shorter pulse duty cycles. We do note however that there may be some exotic pulse shapes for which the standard FFT procedure can still outperform the FFA in terms of sensitivity, if the set of matched filters used in the FFA search does not account for their possible existence. For example, a pulse with two widely spaced identical peaks may see only one of them matched by a set of single-peaked matched filters, resulting in at best $\approx 71\%$ efficiency\footnote{i.e. $1 / \sqrt{2}$, the reason being that unit S/N pulses have unit \textit{square} sum, quite an important detail of this paper}. In that sense, the incoherent harmonic sum could be said to be more agnostic to pulse shape. On the other hand, we have assumed that the fundamental signal frequency falls at the centre of a Fourier frequency bin, i.e. it is an integer multiple of $1/N$. This corresponds to the best-case sensitivity scenario for the FFT method, where all of the signal power is concentrated in Fourier bins with frequencies that are multiples of the fundamental. Otherwise, a significant amount of power leaks to neighbouring bins, a sensitivity-reducing effect known as \textit{scalloping}; it can in principle be eliminated using so-called Fourier interpolation, but not entirely so in practice due to limited computing resources \citep[see \S 4.1 of][for details]{Ransom2002}. In large-scale searches, a computationally cheap version of Fourier interpolation called interbinning is employed, that results in a worst-case loss of signal-to-noise ratio of 7.4\%, which is equivalent to further dividing the FFT's search efficiency by 1.074.

\section{An implementation of the FFA to process large scale pulsar surveys}
\label{sec:implementation}

Implementing an FFA search pipeline that conserves all its desirable properties on ideal data when faced with real-world, imperfect data is challenging. In this section we will describe all the practical problems that need to be surmounted, the proposed solutions that we implemented in our own FFA search software package \textsc{riptide}, and a rationale behind the technical choices that we made. A significant amount of practical knowledge was gained by testing and refining the implementation on a portion of the SUPERB survey \citep{SUPERB}, in particular on observations of known sources that were either faint or made difficult to detect by different strains of radio-frequency interference.

\subsection{Overview of the implementation}

The high-level interface of \textsc{riptide} is written in the \textsc{python} language, while the performance-critical sections of the code are implemented in optimized \textsc{C} functions. \textsc{riptide} provides a library of \textsc{python} functions and classes that can be used to perform interactive analysis and visualisation of dedispersed time series data, using for example \textsc{ipython} or \textsc{jupyter}. The high-level functions of \textsc{riptide} form the building blocks of its pipeline executable, that takes a set of dedispersed time series as input and outputs a number of data products, including a set of files containing the parameters of detected candidate signals and useful diagnostic information (see \S \ref{subsec:pipeline_overview} for details on the pipeline itself). The code is modular so that it can in principle support arbitrary data formats; currently, \textsc{riptide} can process dedispersed time series files produced by either \textsc{presto}\footnote{\url{https://github.com/scottransom/presto}} \citep{Ransom2002} or \textsc{sigproc}\footnote{\url{http://sigproc.sourceforge.net/}} \citep{sigproc} pulsar software packages.

\subsection{Data whitening and normalisation}

In \S \ref{sec:sensitivity} we have demonstrated the optimal sensitivity of the combination of the FFA and matched filtering, but under the significant assumption that the background noise was Gaussian with mean $m$ and standard deviation $\sigma$ known \textit{a priori}, that we took to be 0 and 1 respectively without loss of generality. In practice these parameters have to be estimated from the data directly, in the presence of an additional low-frequency noise component and possibly interference. Devising an efficient and unbiased estimator for $m$ and $\sigma$ in the general case is extremely challenging, as it would require fitting the low-frequency noise component parameters, any detectable signal regardless of its nature, and $m$ and $\sigma$ \textit{simultaneously}. To that we must further add the constraint that evaluating the noise parameters should be done in a small fraction of the time taken by the search process. 

\citet{Cameron2017} have partially avoided this problem by essentially evaluating $m$ and $\sigma$ in every folded profile separately, using its so-called baseline, a region of the pulse that appears to contain no signal defined by empirical criteria. This method has long been employed in pulsar search and timing software, and is implemented for example in the widely used \textsc{psrchive} package \citep{Hotan2004}. However, doing so has a number of drawbacks. In particular, the estimators for the noise parameters have to work with a small number of data points, no larger than the number of phase bins, yielding uncertain estimates that significantly vary from one pulse to the other, and in turn increase the uncertainty of the output S/N values. At longer trial periods, the baseline is also more likely to be contaminated by low-frequency noise. Furthermore, and most importantly, any procedural method to locate the baseline region needs to use criteria such as excluding a putative pulse even where there is none present \citep[e.g. ``Algorithm 2" in][]{Cameron2017}, or search for a contiguous segment of the profile with the lowest mean (implemented in the \textsc{pdmp} utility of \textsc{psrchive}). The two methods above yield estimates of $m$ and $\sigma$ that are systematically biased low in the presence of pure noise, and therefore S/N values that are biased high, which invalidates the optimal sensitivity demonstration made in \S \ref{sec:sensitivity}.

Our priority was therefore to remain within all mathematical assumptions of \S \ref{sec:sensitivity}, at least to the extent allowed by real-world data. We have chosen to directly evaluate the noise parameters from the whole input time series once during pre-processing, in an attempt to obtain unbiased and low-variance estimates. This approach is also used by \citet{Parent2018}. In our implementation, we first remove the low-frequency noise component by subtracting a running median from the input data; the width of the median window is left to the user to choose and optimize. The running median filter has the desirable property of leaving intact any constant impulse whose width is less than half that of the window \citep{GallagherWise1981}. Then the input time series is normalised to zero mean and unit standard deviation before being searched.


\subsection{Searching a period range}
\label{subsec:searching_a_period_range}

\begin{figure*}
\centering
\includegraphics[width=1.00\textwidth]{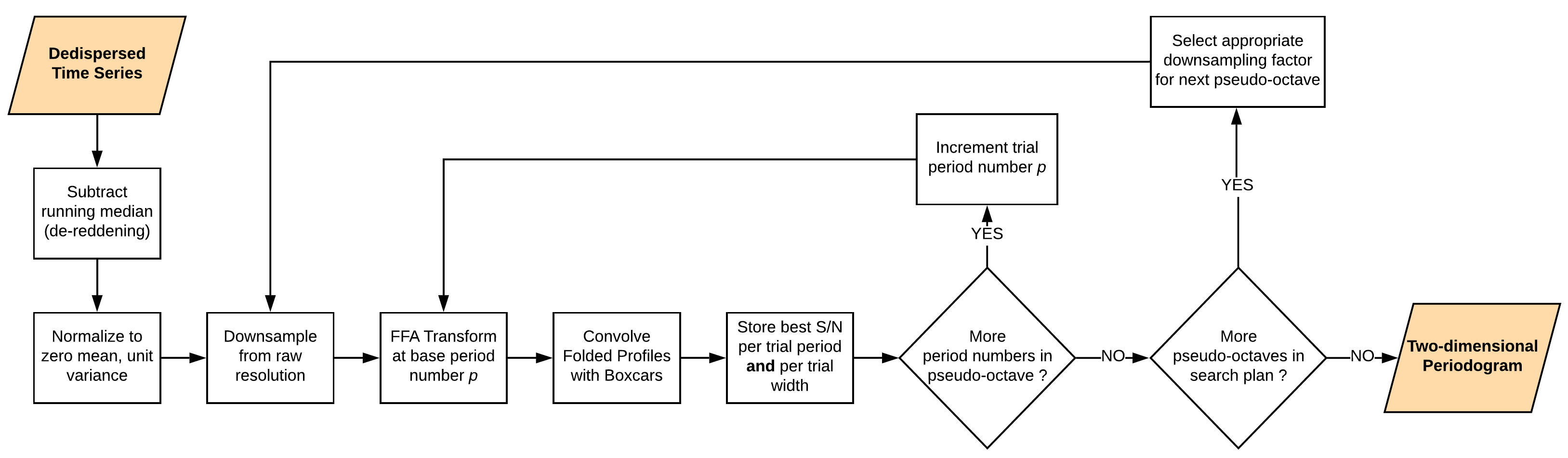}
\caption{Sequence of operations performed by \textsc{riptide} when processing a dedispersed time series over a user-defined period range, see also \S \ref{subsec:searching_a_period_range} for a detailed explanation. The output periodogram is two-dimensional as the code returns the highest signal-to-noise ratio (S/N) per period trial \textit{and} per pulse width trial, rather than just the best S/N per period trial across all pulse widths. In practice this can greatly improve sensitivity to narrow pulses in the presence of interference or low-frequency noise (see text and Fig. \ref{fig:pgram_separation}).}
\label{fig:pgram_flowchart}
\end{figure*}

A given period range $[P_0, P_1]$ could naively be searched by calculating a sequence of folding transforms, with base periods (and therefore number of output phase bins) expressed in number of samples ranging from $p_0 = \floor{P_0 / \tau}$ to $p_1 = \ceil{P_1 / \tau}$, where $\tau$ is the time resolution of the input data, $\floor{}$ and $\ceil{}$ are the floor and ceiling functions respectively. However, if $P_1 / P_0 \gg 1$ this procedure has a highly non-uniform phase and pulse duty cycle resolution, which can become excessively high at trial periods of a few seconds and higher, especially in an era where $\tau$ is typically on order of tens of microseconds. The standard way to circumvent the problem, adopted by previous FFA implementations \citep{Kondratiev2009, Cameron2017}, is to iteratively downsample the input data by a factor of two, and consecutively search so-called \textit{octaves} of the form $[2^k P_0, 2^{k+1} P_0]$, where within each octave the number of phase bins in the output folded profiles increases from $b$ to $2b$ proportionally to the trial period, with $b$ being a user-defined parameter that sets the minimum duty cycle resolution. We have implemented the same procedure in \textsc{riptide}, with one difference worth mentioning: we have provided the option to use real-valued, strictly smaller than 2 downsampling factors which generates more numerous and shorter pseudo-octaves. In practice, the user chooses a minimum and maximum number of output of phase bins $b_\mathrm{min}$ and $b_\mathrm{max}$ respectively, and the data are iteratively downsampled by $f = (b_\mathrm{max} + 1) / b_\mathrm{min}$ rather than $f = 2$. We have made this choice both to maintain the duty cycle resolution much closer to uniform, rather than let it oscillate by a factor of two; furthermore, if what matters most when reporting on a large scale FFA search is the \textit{lowest} duty cycle resolution across the period range explored, then one should attempt to increase the minimum number of phase bins while maintaining the maximum number of phase bins as low as possible to avoid spending additional computing resources. The whole search process is summarized as a diagram in Fig. \ref{fig:pgram_flowchart}.

There are a few of details to be aware of here, firstly that using a \textit{non}-integer factor makes any two consecutive samples in the downsampled data correlated with each other. This is not a problem in principle because when folding, the samples integrated in any given phase bin are not consecutive and therefore uncorrelated. But this may not be the case anymore if chaining (i.e. composing) many downsampling operations, which we carefully avoid doing in our implementation. When moving from one pseudo-octave to the next, we always return to the original highest-resolution data, and downsample it with the appropriately chosen, geometrically increasing factor. The real caveat actually lies in how downsampling by a real-valued factor $f$ affects the noise variance in the output samples. If the input data are uncorrelated Gaussian noise with unit variance, then the sample variance of the output data is $f - \frac{1}{3}$ rather than $f$ as one would expect when using an integer factor. This is shown in \S \ref{subsec:variance_non_integer_downsampling} and properly accounted for in \textsc{riptide}.

\subsection{Matched Filtering}
\label{subsec:matched_filtering}

While the folded profiles can be, and ideally should be, convolved with pulse templates of highly diverse shapes, we have limited ourselves to only boxcars for computing performance reasons at least in the current version of the code. The loss of sensitivity is mild as long as the profile is single-peaked, a loss that has already been evaluated above and in \S \ref{subsec:boxcar_efficiency}. Circular convolution with a set of boxcar filters is particularly fast, as it can be implemented as ``strided differences" of the exclusive prefix sum of the profile being evaluated. If $P$ is the profile array with $n$ bins, its exclusive prefix sum $S$ is an array whose elements are defined as

\begin{equation}
    S_k = \sum_{i=0}^{k-1} P_i,
\end{equation}
with the exception $S_0 = 0$. If $B$ is a boxcar of width $w$ bins and unit height, and $C = P \circledast B$ their circular convolution product, then

\begin{equation}
    C_k = \sum_{i=k}^{k+w-1} P_{i~(\mathrm{mod}~n)} = S_{k+w} - S_k,
\end{equation}
noting that the array index of $P$ is taken modulo $n$ because the convolution is circular, and for the same reason we calculate $S$ over two concatenated repeats of the profile $P$. The height of the boxcar is also properly scaled as a function of the number of pulse periods summed, to follow the unit square sum normalisation of \S \ref{sec:sensitivity}. This implementation is quite efficient on the CPU since it re-uses the profile prefix sum array multiple times, which typically fits into the L1 cache. The sequence of trial boxcar widths used in the search is controlled by two user-defined parameters: a maximum trial duty cycle (20\% by default), and a geometric spacing factor $f_\mathrm{sp}$ between consecutive trial widths ($f_\mathrm{sp} = 1.5$ by default). The first trial width is always $w_0 = 1$ sample, and the whole sequence is generated with the recurrence relation $w_{k+1} = \max \{ \floor{f_\mathrm{sp} w_k}, w_k + 1 \}$ up to a maximum value.

\subsection{Identifying statistically significant candidate signals}

\begin{figure*}
\centering
\includegraphics[width=1.00\textwidth]{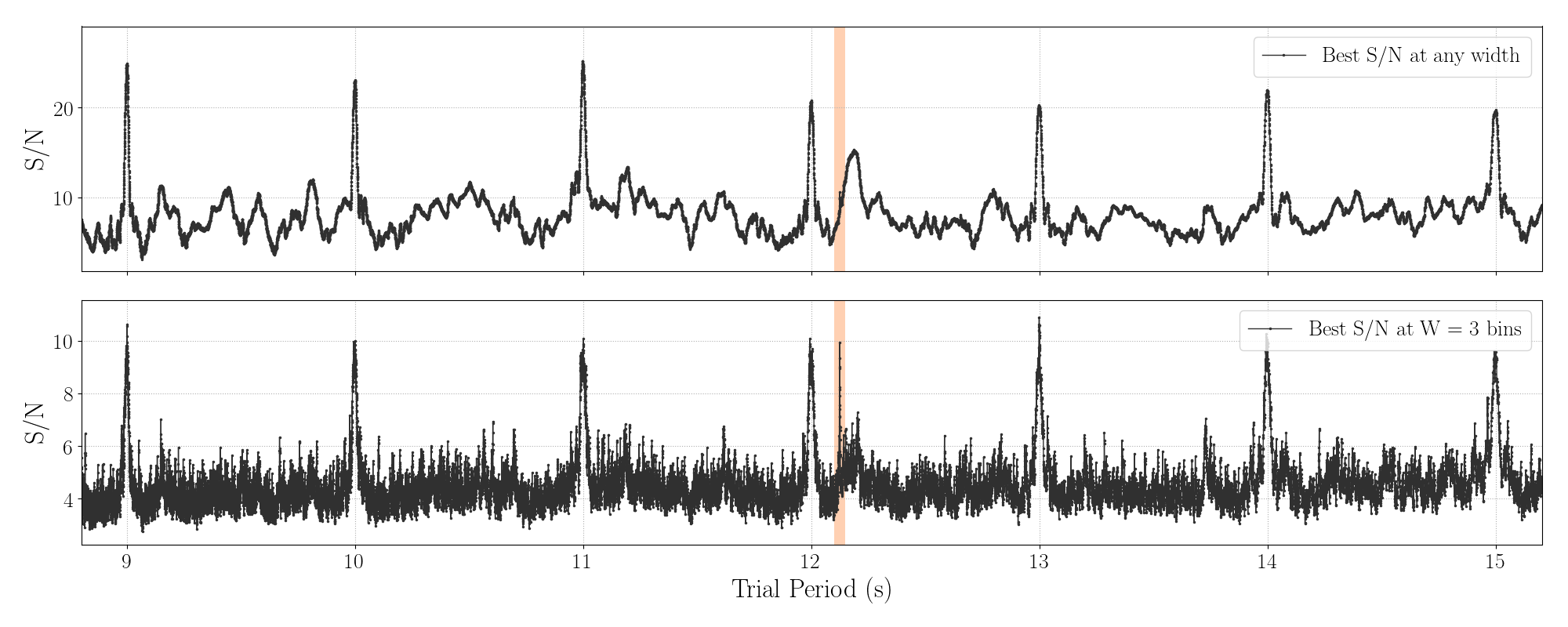}
\caption{Highlighting the advantages of analysing pulse width trials separately when searching for long-period faint sources. Here we have run \textsc{riptide} on a 9-minute Parkes L-band observation of PSR J2251$-$3711, dedispersed at the known dispersion measure of the pulsar. The 12.1-second period of the pulsar is highlighted in orange; many harmonics of an interference source with a 1-second period are detected. Top panel: best S/N obtained across all trial pulse widths as a function of trial period. For most trial periods, the highest S/N is obtained with the widest matched filter due to its optimal response to residual low-frequency noise, and the peak associated with the pulsar is difficult to distinguish from the background. Bottom panel: S/N obtained specifically with the optimal 3-bin width matched filter, where the associated peak could be identified in a blind search. $\textsc{riptide}$ keeps one periodogram per width trial in memory which are then independently searched for peaks, in an attempt to be robust to situations with imperfect data.}
\label{fig:pgram_separation}
\end{figure*}

Overall the combination of FFA and matched filtering returns S/N as a function of trial period, width, and phase, which can be too large to be fully kept in memory especially when processing multiple dispersion measure (DM) trials. FFA implementations typically reduce this output to a so-called periodogram, a one-dimensional array containing the highest S/N as a function of trial period (across all possible trial widths and phases). On ideal data containing only uncorrelated Gaussian noise and a possible periodic signal, identifying statistically significant signal periods is a trivial exercise of applying a threshold to the periodogram; the threshold value is the critical value of the $z$-statistic (Eq. \ref{eq:z_statistic_critval}), i.e. the number of Gaussian sigmas that correspond to the desired false alarm probability. However, situations where this is not applicable arise relatively often in practice. If the data are contaminated by low-frequency noise, then it will manifest itself in \textit{every} folded profile at longer trial periods (typically a few seconds) in the form of a fluctuating baseline, to which the widest matched filters have optimal response. Randomly occurring bursts of interference are also empirically found to have a similar effect. We have noticed on several Parkes L-band test cases that keeping in memory only the best S/N value across all trial widths runs the risk of missing long period pulsar signals that could otherwise be found in a blind search. This is illustrated in Fig. \ref{fig:pgram_separation} on a Parkes multibeam observation of the 12.1-second pulsar PSR J2251$-$3711 \citep{Morello2020} affected by interference: the pulsar emerges as a clear peak if considering the optimal trial width in isolation, but is undetectable if the only information available is the best S/N vs. period plot. We therefore keep in memory the highest S/N as a function of \textit{both} trial period and width, making the output periodograms of \textsc{riptide} two-dimensional; each pulse width trial is then post-processed separately.

\begin{figure*}
\centering
\includegraphics[width=1.00\textwidth]{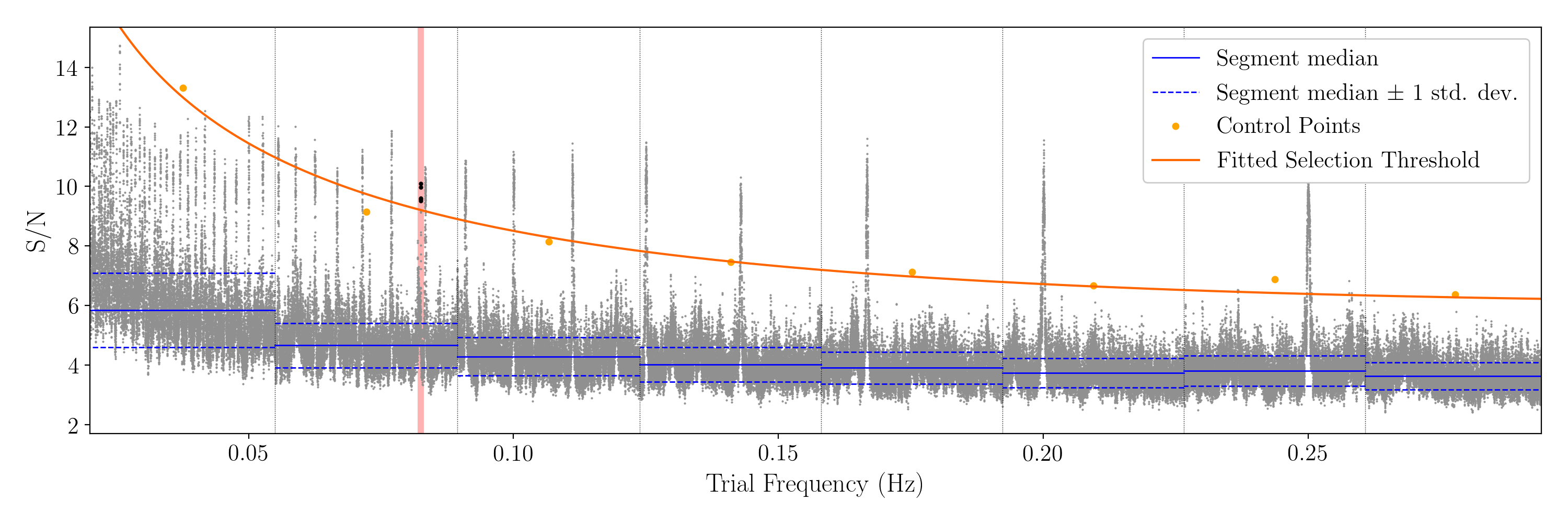}
\caption{An explanation of the peak detection algorithm employed in \textsc{riptide}, using the same observation of PSR J2251$-$3711 as in Fig. \ref{fig:pgram_separation}. For each individual trial width, the S/N vs. trial frequency array is partitioned into equal-sized frequency segments with a width of a few times $1/T$ where $T$ is the length of the input time series. Within every segment, the median $m$ and standard deviation $\sigma$ of the S/N are calculated using robust estimators. Each segment yields a control point with coordinates $(\nu, s)$, where $\nu$ corresponds to the centre frequency of the segment and $s = m + k \sigma$ where $k$ is a parameter set to $k = 6$ by default. Finally, a second-degree polynomial in $\log(\nu)$ is fitted to the control points, yielding the peak selection threshold (orange line). Any points lying above the threshold are clustered into peaks. The detection of the pulsar is highlighted with a red background and bolder points. See text for full details and the design rationale for this algorithm. Note that the frequency segments are much narrower in practice, they have been widened here for readability.}
\label{fig:pgram_fitting_explained}
\end{figure*}

Furthermore, periodograms of contaminated data commonly show a trend of S/N increasing as a function of period, and all reported S/N values above a certain trial period may exceed the theoretical significance threshold. Whether or not a peak is significant can then only be evaluated relative to neighbouring period trials; visual inspection of periodograms is not an option in a large scale survey, and a reliable peak identification algorithm is required. This is a crucial part of a pipeline, which if not carefully designed can vastly reduce the inherent sensitivity of the FFA itself in many imperfect data situations; it should also run in a reasonable fraction of the time required by the computation of the periodogram itself. Our peak finding algorithm is based on evaluating, as a function of trial period, the overall trend of the S/N and the dispersion around it (Fig. \ref{fig:pgram_fitting_explained}), in order to determine how far above the trend a point should lie to be considered significant. We first divide the whole trial frequency range into equal-sized segments. Here we operate on trial frequency rather than trial period, basing ourselves on the property that the complex-valued coefficients of the discrete Fourier transform (DFT) of uncorrelated Gaussian noise are also uncorrelated; the DFT frequency bins are spaced by $1/T$ where $T$ is the integration time, and one can expect the S/N values in an FFA periodogram to decorrelate on a frequency scale on order of $1/T$ as well. The width of the frequency segments is thus chosen to be $\Delta z_{\mathrm{seg}} / T$ where $\Delta z_{\mathrm{seg}}$ is a dimensionless free parameter. In each segment, we calculate the median $m$ of the S/N values and estimate their standard deviation $\sigma$ from the interquartile range (IQR) of the S/N values using $\sigma = \mathrm{IQR} / 1.349$, in order to be robust to outliers. An estimate of the local significance threshold is $s = m + k \sigma$ where $k$ is a dimensionless parameter defining the significance level. All segments thus provide a set of control points ($\nu_i$, $s_i$) where $\nu_i$ is the centre frequency of the $i$-th segment and $s_i$ the estimated local significance threshold. The last step is to fit a function of frequency to this set of points. Here the choice of functional form might aim to reflect that the data are expected to contain two noise components: a Gaussian component with uniform power as a function of frequency, and a low-frequency component whose power is typically modeled as proportional to $\nu^{-\alpha}$ where $\alpha > 0$ is the noise spectral index ($\alpha = 2$ in the case of Brownian noise). With signal power being the square of S/N, a logical choice appears to be a function of the form $s = \sqrt{A \nu^{-\alpha} + B}$; however, we found that it often yields a poor fit to the control points on real test cases, by overestimating the selection threshold at low frequencies. We attribute this to the red noise reduction effect of the running median filter and the presence of interference. Furthermore, this fit is non-linear and does not converge all the time, which is highly undesirable as we aim for maximum reliability. In the end, we found that the best results were obtained with a polynomial in $\log(\nu)$, a function that does not diverge around $\nu = 0$ and can be reliably fit with the linear least-squares method. The degree of the polynomial is exposed as a free parameter, and is equal to 2 by default.

We can note that the exact same candidate selection problems caused by S/N trends are encountered in FFT-based periodicity searches as well, where the low-frequency bins of Fourier spectra contain significant excess power. The classical solution in FFT search implementations is to perform so-called ``spectral whitening", where portions of the power spectra are divided by a factor dependent on the locally estimated power average \citep{IsraelStella1996, Ransom2002}. Our dynamic thresholding algorithm above is conceptually similar.

\subsection{Improving performance with dynamic duty cycle resolution}

A useful feature of our implementation is the option to split the search period range into a number of user-defined segments, each using a different number of phase bins, where the idea is to use more bins (i.e. a higher duty cycle resolution) at longer trial periods. Indeed the complexity of an FFA search scales unfavourably with both the minimum period being searched and the maximum desired duty cycle resolution. Searching an octave means performing a sequence of folding transforms and convolving every output profile with matched filters. Using notations from \S \ref{sec:algorithm}, we note $N$ is the total number of samples in the input time series after any appropriate downsampling, $p$ the base period in samples of the folding transform and $m$ the number of rows such that $N = mp$. The cost of the folding transform itself is $m p \log_2(m)$ additions if $m$ is a power of two \citep{Staelin1969}. The cost of matched filtering with boxcars, using the method of \S \ref{subsec:matched_filtering}, is equal to the product of the number of rows, number of phase bins ($p$) and number of boxcar matched filters. If geometrically spacing the consecutive pulse width trials by a factor of 2, the number of matched filters is approximately equal to $\log_2(p)$. The number of arithmetic operations associated with a single folding transform can thus be estimated by

\begin{equation}
    \mc{C}_{\rm{FT}} = mp \left( \log_2(m) + \log_2(p) \right) = N \log_2(N)
\end{equation}
Searching an octave where the trial periods range between $b$ and $2 b$ therefore costs $\mc{C}_{\rm{octave}} = b N \log_2(N)$. Here $b$ represents the shortest trial period in number of samples and determines the \textit{minimum} duty cycle resolution $1/b$. The asymptotic complexity of searching several octaves using iterative downsampling is dominated by the cost of the first, shortest period octave; the cost of searching a time series is thus $\mc{C}_{\rm{TS}} = \mc{O}(b N \log_2(N))$. If we note $T$ the integration time and $P_0$ the shortest trial period in seconds, then the input data can be initially downsampled to an optimal time resolution of $\tau = P_0 / b$, which means the effective number of samples to search is $N = T / \tau$. Noting $\nu_0 = 1 / P_0$ the highest trial frequency, we can re-express $\mc{C}_{\rm{TS}}$ as

\begin{equation}
\label{eq:complexity_search_timeseries}
    \mc{C}_{\rm{TS}} = \mc{O}(T \nu_0 b^2 \log(T \nu_0 b)).
\end{equation}
However, if searching a range of dispersion measures (DMs), one must limit the pulse broadening effects of dedispersing at an incorrect DM \citep[e.g.][]{CordesMcLaughlin2003}. To achieve adequate DM space coverage, the trial DM step needs to be proportional to the minimum pulse width being searched $W_\mathrm{min} = P_0 / b$, which means that the number of required DM trials scales with $\nu_0 b$. The total asymptotic complexity of searching all dispersion measures is thus

\begin{equation}
    \mc{C}_{\rm{total}} = \mc{O}(T \nu_0^2 b^3 \log(T \nu_0 b)).
\end{equation}
The $b^3$ term is a strong incentive to avoid using the same duty cycle resolution across a search period range that may span two orders of magnitude or more in practice. When using the full pipeline, \textsc{riptide} users are therefore given the option to partition the period search range into any number of arbitrary sub-ranges, each with the duty cycle resolution of their choice depending on computing resources available. This might enable for example searching periods down to 100 ms or even shorter with a small ($\approx 100$) number of phase bins, while keeping a high duty cycle resolution ($\gtrsim 1,000$ phase bins) for periods of a second or longer.

\subsection{Full pipeline overview}
\label{subsec:pipeline_overview}

In a large-scale pulsar survey processing scenario, an FFT-based search that would be running simultaneously already requires an observation to be dedispersed into a set of time series each with their trial dispersion measure, with a DM step almost certainly finer than required by an FFA search. The \textsc{riptide} pipeline is designed to take advantage of this situation, taking a set of existing DM trials as an input and processing only the minimal subset necessary to achieve adequate DM space coverage; the trial DM step is chosen using the classical method, that is based on limiting the pulse broadening time scale due to dedispersing the data at a DM different from that of the source \citep[e.g. Chapter 6 of][]{PulsarHandbook}. If running \textsc{riptide} on its own, then the user is responsible for dedispersing the data with the appropriate DM step beforehand. All selected DM trials are then distributed between available CPUs using the \texttt{multiprocessing} standard library of \textsc{python}, and processed in parallel. For each DM trial a periodogram is produced and searched for peaks, but only the peak parameters are accumulated in memory. Once all time series have been processed, the parallel section of the pipeline ends. All reported peaks are then grouped into clusters in frequency space, and clusters that are harmonics of others are flagged. The user decides whether such harmonics are kept and later output as candidates, or discarded immediately; finally, for each remaining cluster, a candidate object is produced. The code saves the following data products:

\begin{enumerate}
    \item A table of all detected peak parameters \textit{before} clustering. These include: period, DM, width in phase bins, duty cycle, S/N.
    \item A table of cluster parameters, with any harmonic relationships. The cluster parameters are simply that of its brightest member peak object.
    \item Candidate files in JSON format each containing: a list of best-fit parameters (period, DM, S/N, duty cycle), a sub-integrations plot produced by folding the de-reddened and normalised time series associated with the best-fit DM, all available metadata of the observation (e.g. coordinates, integration time, observing frequency, etc.), and a table of all detected peaks associated with the candidate.
    \item Candidate plots, although this can be disabled to save time; plots can still be produced later.
\end{enumerate}

Saving intermediate data products enables one to test all post-processing steps on known pulsar observations, and further improve them in the future. It is important to verify for example if a known pulsar is missed because it is genuinely undetectable, or if it registers as a peak and then is lost due to erroneous clustering or harmonic flagging.

\subsection{Benchmarks}

\begin{figure*}
\centering
\includegraphics[width=1.00\textwidth]{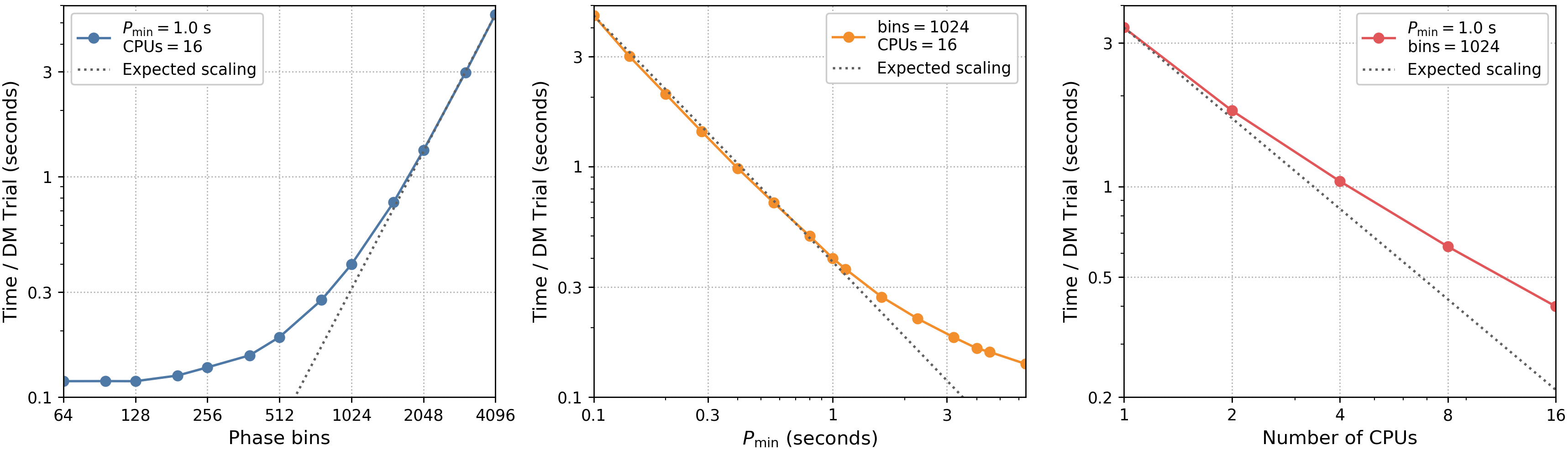}
\caption{Processing time per DM trial of the \textsc{riptide} pipeline, evaluated on a set of 128 artificial DM trials with a length of 9 minutes and a sampling time of 64 $\mu$s. The pipeline was initially configured to search trial periods from 1 to 120 seconds with 1024 phase bins, and allowed to use 16 CPU cores. We then measured how changing one configuration parameter at a time from this initial setup affected the execution time. The plots show the total processing time per DM trial as a function of the average number of output phase bins (left), minimum trial period (middle) and number of CPU cores (right). For the left and middle plots, the expected scaling is extrapolated from the most intensive test case (i.e. with the highest run time) using Eq. \ref{eq:complexity_search_timeseries}, which gives the asymptotic complexity of searching a single time series. For the right plot, the expected scaling is a linear performance increase as a function of number of CPU cores. See text for more details on the experimental setup and a discussion of the results.}
\label{fig:benchmarks}
\end{figure*}

To measure the speed of the \textsc{riptide} pipeline, we have decided to avoid using a pulsar observation, because the total number of candidate files that must be produced by the final stage of the pipeline is dependent on the signal content of the data. For reproducibility purposes, we have therefore measured the execution time of the \textsc{riptide} pipeline (version 0.1.0) on a set of 128 DM trials made of artificially generated Gaussian noise. Each time series was 9 minutes long, containing $2^{23}$ time samples with a $64 \mu$s sampling interval, reproducing the parameters of recent Parkes multibeam receiver surveys \citep{Keith2010, SUPERB}. The benchmark was conducted on a node of the OzStar computing facility, reserving 16 Intel Gold 6140 @ 2.3GHz CPU cores to search the data in parallel. Here we deliberately did \textit{not} use the dynamic duty cycle resolution feature previously described, choosing instead to specify a single, wide period search range with the same $b_\mathrm{min}$ and $b_\mathrm{max}$ parameters in order to keep the experiment simple. We provided \textsc{riptide} with the following base configuration:
\begin{itemize}
    \item Minimum search period $P_\mathrm{min} = 1.0$ s
    \item Maximum search period $P_\mathrm{min} = 120.0$ s
    \item Average number of phase bins $b_\mathrm{avg} = 1024$, for a duty cycle resolution of $\delta = 0.1\%$. In practice we specified the code parameters $b_\mathrm{max}$ and $b_\mathrm{min}$ to be equal to $b_\mathrm{avg}$ plus and minus 4 percent, respectively.
    \item $N_\mathrm{proc} = 16$ parallel processes, one per available CPU core
\end{itemize}
We then measured how individually changing the parameters $P_\mathrm{min}$, $b_\mathrm{avg}$ and $N_\mathrm{proc}$ (while keeping all other parameters fixed) affected the total processing time per DM trial. The results are shown in Fig. \ref{fig:benchmarks}. The execution times as a function of minimum trial period and number of phase bins are well predicted by the asymptotic complexity expression given in Eq. \ref{eq:complexity_search_timeseries}. A deviation from the prediction is observed where the search is computationally cheap, that is for both lower number of phase bins and higher minimum trial periods: in this regime, the total execution time is instead dominated by the overheads of reading, de-reddening and normalising the input data. However, the performance of the pipeline was not found to scale quite linearly with the number of CPUs employed; although some operations of the pipeline run sequentially and do not benefit from multiple CPUs, we found that the non-parallel sections of the code consumed a negligible amount of time in this experiment. We therefore attribute the observed effect to limited memory bandwidth, for which multiple CPUs likely compete. Overall, a single 9-minute DM trial can be processed in approximately one second on a modern compute node with 16 CPU cores, while using at least 1,000 phase bins for trial periods of a second or longer.

\section{Discussion}
\label{sec:discussion}

\subsection{A more realistic radiometer equation to determine detectable flux densities}

The following form of the radiometer equation \citep[][]{Dewey1985, PulsarHandbook} has often been used to evaluate the minimum detectable average flux density $S_\mathrm{min}$ in a pulsar survey:

\begin{equation}
    S_\mathrm{min} =  \beta  \frac{\mathrm{(S/N_{min})} T_\mathrm{sys}}{G \sqrt{n_\mathrm{p}~t_\mathrm{int}~\Delta f}} \sqrt{\frac{W}{P-W}},
\end{equation}
where $\beta$ is a loss factor due to digitization\footnote{although some authors include other losses in $\beta$, which is relevant to the discussion here, see below.}, $T_\mathrm{sys}$ is the system temperature, $G$ the telescope gain, $n_p$ the number of polarization channels summed, $t_\mathrm{int}$ the integration time, $\Delta f$ the observing bandwidth, $P$ the signal period, and $W$ the effective pulse width in the dedispersed data, i.e. with \textit{all} broadening effects taken into account. $\mathrm{S/N_{min}}$ is quoted as the ``detection threshold" in the original paper, which is incorrect; instead one should substitute it for the actual minimum detectable amplitude, the \textit{upper limit} of the search. Indeed the detection threshold is the number of Gaussian sigmas that corresponds to the false alarm probability $\alpha$, i.e. the critical value of the Z-statistic (Eq. \ref{eq:z_statistic_critval}); the upper limit is equal to the detection threshold if and only if performing phase-coherent folding followed by matched filtering with the exact pulse shape of the signal (Eq. \ref{eq:z_statistic_upperlim}). \citet{Yu2018} has previously reported the discrepancy, and that the equation above is misused in a number of pulsar survey papers. If we retain $\mathrm{S/N_{min}}$ as the \textit{detection threshold} in the equation above, then the search efficiency $\mc{E}$ needs to be taken into account as follows:
\begin{equation}\label{eqn:radiometer_fixed}
    S_\mathrm{min} =  \beta  \frac{\mathrm{(S/N_{min})} T_\mathrm{sys}}{\mc{E} G \sqrt{n_\mathrm{p}~t_\mathrm{int}~\Delta f}} \sqrt{\frac{W}{P-W}}.
\end{equation}
For an FFA search, $\mc{E}_\mathrm{FFA} = 0.93$ can be used as long as the duty cycles considered are resolved in the folded output. For an incoherent FFT search, $\mc{E}$ is a function of effective pulse duty cycle $\delta = W/P$; for practical use, we have fitted a sigmoid function to the search efficiency shown in the right panel of Fig. \ref{fig:sensitivity_comparison}. Assuming that up to 32 harmonics can be summed:

\begin{equation}
    \mc{E}_\mathrm{FFT}(\delta) \approx \left( 1 + 4.73 \times 10^{-2} ~\delta^{-0.627} \right)^{-1},
\end{equation}
which approximates the exact value derived in \S \ref{sec:sensitivity_fft} with a relative error no larger than 6\% for $10^{-4} \leq \delta \leq 0.5$. The exact value for the median pulsar duty cycle of 2.8\% \citep[from the ATNF pulsar catalogue,][]{PSRCAT} is $\mc{E}_\mathrm{FFT} = 0.70$, which remains valid even if summing a maximum of 16 harmonics. These corrections do not account for any additional reduction in sensitivity incurred from data imperfections, various software implementation details and the possibilities of losing a possible detection in the many stages of the search process. Whenever possible, the true minimum detectable average flux density should be evaluated via signal injection in real data processing conditions, as previously done in \citet{Lazarus2015} and \citet{Parent2018}. Finally, we note that there are multiple conventions as to what the $\beta$ factor represents in the radiometer equation above. As an example, \citet{Johnston1992} have included in $\beta$ not only losses due to digitization, but also what they called ``approximations made in the software"; it could potentially refer to an estimate of the search efficiency. Likewise, \citet{Dewey1985} defined $\beta$ as ``various system losses".

Above and beyond assessing the \textit{detection} efficiency, a further caveat is needed when using values calculated by Eqn.~\ref{eqn:radiometer_fixed} to infer intrinsic pulsar properties. It would be incorrect to interpret flux density values determined in this way, that apply to detection thresholds in the final dedispersed data stream, as simply-scaled versions of intrinsic pulsar properties. For example it is common to scale up flux densities determined from the (even uncorrected) radiometer equation by the square of the distance to a pulsar to infer an intrinsic property. When doing this, at the very least an extra efficiency factor should come into play to account for the various transfer functions of the interstellar medium, telescope system, dedispersion procedures, etc., and in reality the inversion might be quite complex and line-of-sight dependent. The end result, as to what fraction of the intrinsic population is probed by these searches, contains this difficult-to-account-for incompleteness.

\subsection{Improving FFT searches with coherent harmonic summing}

\begin{figure}
\centering
\includegraphics[width=1.00\columnwidth]{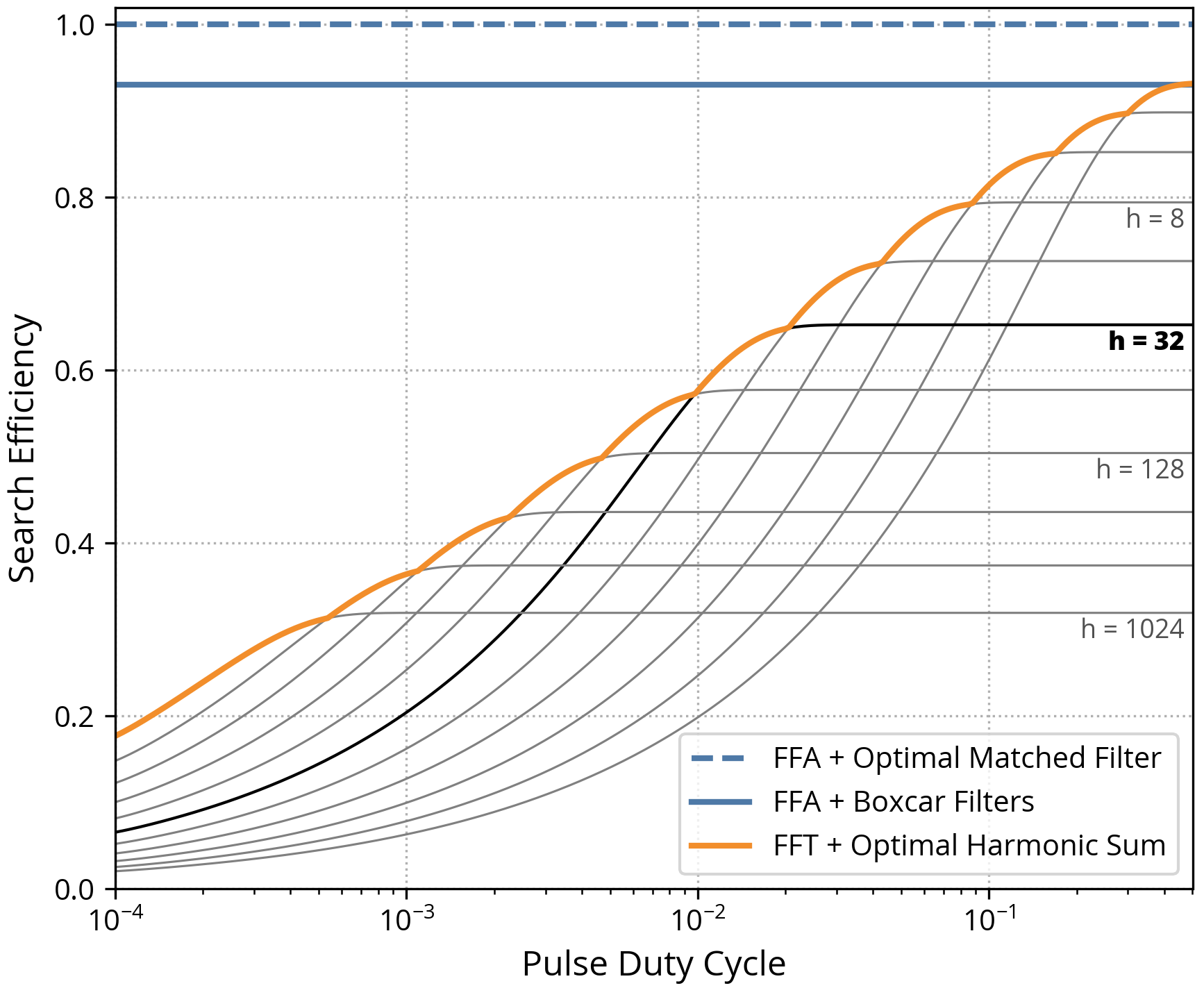}
\caption{Same search efficiency plot as in Fig. \ref{fig:sensitivity_comparison}, also assuming a 7-$\sigma$ significance level and a Gaussian pulse, except that we have allowed for the incoherent summation of up to 1024 harmonics. We have extended the X-axis to include pulse duty cycles down to 0.01\%. The usual limit of 32 harmonics summed in most search codes is highlighted in black. Summing more harmonics is beneficial for duty cycles shorter than about 1\%, but cannot compensate for the lack of phase-coherence of this search procedure.}
\label{fig:sensitivity_comparison_1024h}
\end{figure}

It has been previously suggested \citep{Kondratiev2009, Lazarus2015, Parent2018} that the reduced sensitivity to narrow pulses experienced by the standard FFT procedure is foremost caused by the limit of 32 harmonics usually imposed on the incoherent harmonic sums. We have put this idea to the test in Fig. \ref{fig:sensitivity_comparison_1024h}, running the sensitivity comparison again but this time allowing up to 1024 harmonics to be summed. From the results we instead conclude that the lack of sensitivity of the standard FFT+IHS method to narrower pulses is mainly a consequence of its incoherence, i.e. the discarding of all phase information contained in complex-valued Fourier spectra. It is possible however to achieve the same level of sensitivity as the coherent FFA-based procedure in a Fourier domain search. Indeed the Z-statistic can be calculated as follows, remarking that the unitary DFT conserves the complex dot product:

\begin{equation}
    Z(\mb{x}) = \mb{X} \cdot \mb{S} =  \sum_{i=0}^{N-1} X_i \overline{S_i},
\end{equation}
where $\mb{X}$ and $\mb{S}$ are respectively the unitary DFTs of the data and the signal template being searched for, and the overline represents complex conjugation. The formula above represents a coherent form of harmonic summing, which is equivalent to folding and matched filtering in the time domain. Coherent harmonic summing is a potentially attractive solution, as it could be integrated into an existing FFT+IHS pipeline where the Fourier transform is already performed for free, and used as a complementary method to cover the short duty cycle regime where the incoherent search is least efficient. If too costly, this coherent search could be limited to either longer signal periods or isolated pulsars, ignoring any Doppler-induced variations of signal period due to binary motion which are expensive to compensate for \citep{JK91, Ransom2002}.

\subsection{Efficacy of the FFA in a real search}

We have covered in-depth the topic of the theoretical sensitivity of the FFA, but its ability to discover new pulsars in practice is of high importance. \textsc{riptide} has successfully been used to discover new pulsars that were missed by FFT search codes running on the same data. An early development version of \textsc{riptide} was used to process a portion of the SUPERB survey \citep{SUPERB}. This search was highly experimental for several reasons; the pipeline used a crude peak detection algorithm to analyze periodograms which may have resulted in missed detections, and employed a low number of phase bins (250) even at the longest search periods which clearly limited its ability to find short duty cycle pulsars similar to the 23.5-second PSR J0250$+$5854 \citep{Tan2018}. Furthermore, the accuracy of the candidate selection algorithm employed was limited compared to the one running on the candidates produced by the standard FFT search. In spite of these limitations, 9 previously unknown pulsars missed in the FFT analysis were discovered, with periods ranging from 373 ms to 2.4 seconds and duty cycles larger than one percent. Timing solutions for these discoveries are presented in \citet{Spiewak2020}. A new, deeper FFA search of the entire SUPERB survey and of several other Parkes archival surveys will be performed soon (Morello et al., in prep.).

More recently, \textsc{riptide} has been used to discover the previously unknown pulsar PSR J0043$-$73, with a 937 ms period, in a deep targeted search of the Small Magellanic Cloud \citep{Titus2019}. This source was not detected by the FFT-based \textsc{presto} running on the same data, although it should be noted that to save computing time, it was only allowed to sum up to 8 harmonics. In the same data, the known pulsar PSR J0131$-$7310, with a 348 ms period, was also detected exclusively by \textsc{riptide}. Furthermore, a complete re-processing of the LOFAR Tied-Array All-Sky Survey \citep[][]{Sanidas2019} is currently in progress where \textsc{riptide} is being run in parallel to \textsc{presto}. The results will be reported in a future paper (Bezuidenhout et al., in prep.).

The fact that \textsc{riptide} has discovered a number of pulsars with relatively short periods (less than a second), in a regime where the FFA was previously thought to bring no substantial benefits, is both encouraging and consistent with the theoretical sensitivity analysis presented here. Any difference in the number of pulsars detected between FFA and FFT search codes may result not only from the higher intrinsic sensitivity of the FFA, that we have demonstrated in this paper, but also from other implementation details such as low-frequency noise and interference mitigation schemes, which may be of equal if not higher importance. A thorough empirical comparison between \textsc{riptide} and other FFT search codes (\textsc{presto} in particular) on real data will be performed in the aforementioned upcoming publications.

\section{Conclusion}
\label{sec:conclusion}

We have analytically shown that a search procedure that combines a Fast Folding Algorithm with matched filtering is optimally sensitive to periodic signals in the presence of uncorrelated Gaussian noise. In particular it is more sensitive than the standard FFT technique with incoherent harmonic summing to \textit{all} periodic signals. We have defined a search efficiency term, to rigorously compare the sensitivities of pulsar searching procedures; the optimal search efficiency is 1 when phase-coherently folding the input data and correlating the output with a matched filter reproducing the exact shape of the signal. We have estimated the practical efficiency of both an FFA search (0.93), and of an FFT search with incoherent harmonic summing (0.70 for the median pulsar duty cycle). The efficiency of the FFT+IHS is even lower for shorter pulse duty cycles $\delta$. For example it is a factor of at least 4 less sensitive than the FFA for $\delta = 0.1\%$. The advantage of the FFA could be even greater in practice on long-period signals due to how FFT search codes mitigate the effects of low-frequency noise, although this must be investigated further.

There are two main consequences. Firstly, that a portion of the pulsar phase space has not been properly searched in nearly all radio pulsar surveys, and that the inclusion of an FFA search code in all processing pipelines should now become standard practice. For that purpose, we have made our FFA implementation, \textsc{riptide}, publicly available. Secondly, we have discussed how the search efficiency term must be properly taken into account when trying to infer properties of the general pulsar population from the results of a pulsar search survey; we have warned against a potential misuse of the radiometer equation where the search efficiency is implicitly assumed to be equal to 1, even for an incoherent and thus less efficient search method. This leads to a significant overestimate of the sensitivity of a survey, and could have important consequences for pulsar population synthesis studies, which may also need to take into account the duty cycle dependence of the standard FFT+IHS method's efficiency.

\section*{Data Availability}

The discovery observation of PSR J2251$-$3711, used to demonstrate the peak finding algorithm implemented in \textsc{riptide} (Figs. \ref{fig:pgram_separation} and \ref{fig:pgram_fitting_explained}), as well as the benchmark data (Fig. \ref{fig:benchmarks}) will be shared on reasonable request to the corresponding author.

\section*{Acknowledgements}

VM and BWS acknowledge funding from the European Research Council (ERC) under the European Union's Horizon 2020 research and innovation programme (grant agreement No. 694745). Code benchmarks were performed on the OzSTAR national facility at Swinburne University of Technology. OzSTAR is funded by Swinburne and the National Collaborative Research Infrastructure Strategy (NCRIS). VM thanks Greg Ashton, Scott Ransom, Anne Archibald and Kendrick Smith for useful discussions, and the organizers of the IAU symposium ``Pulsar Astrophysics - The Next 50 Years" for making some of these discussions possible in the first place. Finally, we thank the anonymous referee for helping considerably improve the clarity of the paper.




\bibliographystyle{mnras}
\bibliography{references} 

\appendix

\section{Mathematical derivations}

\subsection{The likelihood-ratio test for the presence of a signal with known shape}
\label{subsec:lrt_expression}

With the context of the hypothesis testing problem of section \ref{sec:sensitivity}, let us write the log-likelihood function of the data $\mb{x}$. Recall that we have modelled the data as the sum of: a known signal $\mb{s}$ normalised to unit square sum scaled by an amplitude parameter $a$, and of independent samples of normally distributed noise $\mb{w}$ (Eq. \ref{eq:input_data}). The data $\mb{x}$ follow a multivariate Gaussian distribution with:
\begin{itemize}
    \item mean $\mb{\mu} = a \mb{s}$ 
    \item covariance matrix that is equal to the identity matrix, because the noise samples $w_i$ are assumed to be mutually independent, and thus uncorrelated
\end{itemize}
This means that the log-likelihood function of $a$ given the data $\mb{x}$ is

\begin{equation}
\label{eq:log_likelihood_expansion}
\begin{split}
    \ln \mc{L}(a|\mb{x}) &= -\frac{N}{2} \ln({2\pi}) - \frac{1}{2} \norm{\mb{x} - a \mb{s}}^2 \\
                         &= -\frac{N}{2} \ln({2\pi}) - \frac{1}{2} (\norm{\mb{x}}^2 + a^2 - 2 a \mb{x} \cdot \mb{s}),
\end{split}
\end{equation}
where $N$ is the length of vector $\mb{x}$, and we used the fact that $\norm{\mb{s}}^2 = 1$. We then apply the definition of the likelihood-ratio test (Eq. \ref{eq:likelihood_ratio_test_definition}) to our hypothesis testing problem, recalling that the null hypothesis is $a = 0$:

\begin{equation}
\begin{split}
\Lambda(\mb{x}) = \frac{ \mc{L}(0|\mb{x}) }{ \sup_{a \geq 0} \mc{L}(a |\mb{x}) }
\end{split}
\end{equation}
Differentiating Eq. \ref{eq:log_likelihood_expansion} with respect to $a$ shows that $\mc{L}(a|\mb{x})$ reaches its maximum value for $a^* = \mb{x} \cdot \mb{s}$. This also means that the test statistic $Z(\mb{x}) = \mb{x} \cdot \mb{s}$ is, by definition, the maximum-likelihood estimator for the parameter $a$. From there, the compact expression of Eq. \ref{eq:likelihood_ratio_test} for $\ln \Lambda(\mb{x})$ follows.

\subsection{Optimality of the Z-statistic}
\label{subsec:z_statistic_optimality_demo}

Here we will show using the Karlin-Rubin theorem, as formulated in \citet{StatisticalInference}, that a test based on the Z-statistic is the most powerful for the presence of the signal $\mb{s}$ in the data regardless of significance level $\alpha$. Recall that our two hypotheses involving the signal amplitude parameter $a$ are $H_0: a = 0$ and $H_1: a>0$. The theorem requires the following two conditions to be met:
\begin{enumerate}
    \item $Z$ is a \textit{sufficient statistic} for the parameter $a$; this means the test statistic $Z$ captures all the information that the data $\mb{x}$ can provide about the parameter $a$.
    \item $Z$ has a monotone likelihood ratio; that is, if $g(z|a)$ is the probability distribution function (PDF) of $Z$, then for any $a_2 > a_1$ the ratio $g(z|a_2) / g(z|a_1)$ is a monotone function of $z$. Intuitively, this expresses the idea that the larger the test statistic $Z$, the more likely the alternative hypothesis is to be true.
\end{enumerate}
Under the conditions above the theorem states that the following test is most powerful at significance level $\alpha$:
\begin{itemize}
    \item Choose $H_1$ if $Z \geq \eta$ with $\eta$ such that $P(Z \geq \eta | H_0) = \alpha$. The value of $\eta$ is given by Eq. \ref{eq:z_statistic_critval}.
    \item Choose $H_0$ otherwise
\end{itemize}
We can show that both conditions above are indeed met. Firstly, $Z$ is a sufficient statistic for $a$ by application of the Fisher-Neyman factorization theorem \citep[e.g. Chapter 6 of][]{StatisticalInference}; the joint PDF of the data $f(\mb{x}|a)$ can be expressed as follows, knowing that the data follow the multivariate Gaussian distribution stated in \S \ref{subsec:lrt_expression}:

\begin{equation}
    f(\mb{x}|a) = \frac{1}{\sqrt{(2 \pi)^N}}   \exp \left( -\frac{1}{2} \norm{\mb{x}}^2 \right)   \exp \left( a Z(\mb{x}) - \frac{1}{2} a^2 \right).
\end{equation}
Here $f(\mb{x}|a)$ is the product of two terms: one that depends on the data but \textit{not} on the parameter $a$, and another (the rightmost exponential) that does depend on the parameter $a$, but only depends on the data through the Z statistic $Z(\mb{x}) = \mb{x} \cdot \mb{s}$. The factorisation theorem thus applies and $Z$ is a sufficient statistic for $a$. Secondly, we have previously shown (Eq. \ref{eq:z_statistic_expansion}) that $Z$ follows a normal distribution with mean $a$ and unit variance, and thus the PDF of $Z$ is given by $g(z|a) = \exp ( -(z-a)^2 / 2 ) / \sqrt{2 \pi}$. For any $a_2 > a_1$ we have

\begin{equation}
    2 \ln \frac{g(z|a_2)}{g(z|a_1)} = 2(a_2 - a_1)z + (a_1^2 - a_2^2),
\end{equation}
which shows that $g(z|a_2) / g(z|a_1)$ is a monotone function of $z$, completing the demonstration.

\subsection{Efficiency of boxcar matched filters in a blind search for Gaussian pulses}
\label{subsec:boxcar_efficiency}

Although in a practical FFA search the folded profiles are discretely sampled, we will assume that the width of a phase bin is much smaller than the width of the Gaussian-shaped pulse being searched for. This enables us to work with continuous functions instead and derive analytical expressions for the search efficiency when using a family of boxcar pulse templates. We model the true pulse shape by a Gaussian function of the form
\begin{equation}
    g(\phi) = A \exp(-\frac{\phi^2}{2 \sigma^2}),
\end{equation}
where $\phi$ represents pulse phase in units of the pulse period, $\sigma$ the Gaussian's standard deviation in the same units, and $\phi \in [-1/2, +1/2]$. $A$ is a normalisation factor to be chosen so that, in line with the convention of \S \ref{sec:sensitivity}, the square sum of $g$ over one pulse period is equal to 1. This is written as
\begin{equation}
    \int_{-1/2}^{1/2} A^2 \exp(-\frac{\phi^2}{\sigma^2}) d\phi = 1.
\end{equation}
If we further assume $\sigma \ll 1$, then the function is essentially zero outside the integration interval, and replacing the integration bounds by $-\infty$ and $+\infty$ does not change the value of the integral. Doing so in the equation above then yields $A^2 \sqrt{\pi \sigma^2} = 1$, and our properly normalised Gaussian pulse is therefore

\begin{equation}
\label{eq:gaussian}
    g(\phi) = \left( \pi \sigma^2 \right)^{-1/4} \exp(-\frac{\phi^2}{2 \sigma^2}).
\end{equation}
Our search template is a boxcar filter of width $w$ also expressed in units of the pulse period
\begin{equation}
  t(\phi) =
  \begin{cases}
    1/\sqrt{w} & \text{if $\phi \in [-w/2, +w/2]$} \\
    0 & \text{otherwise}
  \end{cases}
\end{equation}
where the unit square sum normalisation has also been applied. The efficiency $\mc{E}$ (Eq. \ref{eq:efficiency}) is equal to the scalar product of $g$ with $t$, which is a function of $w$:
\begin{equation}
\begin{split}
    \mc{E}(w) &= \int_{-w/2}^{+w/2} g(\phi) t(\phi) d\phi \\
         &= \pi^{-1/4} \sigma^{-1/2} w^{-1/2} \int_{-w/2}^{+w/2} \exp(-\frac{\phi^2}{2 \sigma^2}) d\phi \\
         &= \left( \frac{4 \pi \sigma^2}{w^2} \right)^{1/4} \mathrm{erf} \left( \frac{w}{2 \sigma \sqrt{2}} \right)
\end{split}
\end{equation}
where the integration range has been limited to $[-w/2, +w/2]$ since $t$ is uniformly zero outside of these bounds. The integral in the expression above has then been expressed in terms of the error function $\mathrm{erf}$. For increased clarity, let us define the half-width of the boxcar in units of sigma $u = \frac{w}{2 \sigma}$. We can then write the more compact expression

\begin{equation}
    \mc{E}(u) = \left( \frac{\pi}{u^2} \right)^{1/4} \mathrm{erf} \left( \frac{u}{\sqrt{2}} \right).
\end{equation}
We find numerically that $\mc{E}$ reaches its maximum for $u^* = 1.40$ and that $\mc{E}(u^*) = 0.943$. In other words, the boxcar filter that optimally matches a Gaussian function with standard deviation $\sigma$ has a width $w^* = 2.80\sigma = 1.19~\rm{FWHM}$ and recovers 94.3\% of the true signal-to-noise ratio of the pulse. In a blind search, $\sigma$ is unknown however; a good strategy is to try a one-dimensional grid of boxcar widths that are powers-of-two multiples of some minimum value (one phase bin in practice). In this case, any interval of the form $[w, 2w]$ is guaranteed to contain a grid point, and in particular $[w^*/\sqrt{2}, w^* \sqrt{2}]$ is not only centered (in log-scale) on the optimum $w^*$ but also always contains the grid point closest (in log-scale as well) to $w^*$. A good estimate of the average efficiency of the search $\bar{\mc{E}}$ is therefore:

\begin{equation}
    \bar{\mc{E}} = \frac{\sqrt{2}}{u^*} \int_{u^* / \sqrt{2}}^{u^* \sqrt{2}} \mc{E}(u) du \approx 0.930,
\end{equation}
where the leading factor $\frac{\sqrt{2}}{u^*}$ is the inverse of the length of the integration interval. The worst-case scenario is reached for $u = u^* \sqrt{2}$, where $\mc{E} = 0.901$. Two major caveats to keep in mind here are that the search efficiency may be significantly lower for pulses with multiple peaks, and that the above derivation was done by approximating discrete folded pulse profiles and templates with continuous functions, which is valid only if the number of phase bins is large enough for the pulse to be well-resolved.

We note that the above derivation is directly applicable to single-pulse searches as well. Our 0.943 peak efficiency factor differs from the value of 0.868 given in the appendix of \citet{McLaughlin2003}; although they do not provide details about how their value is derived, the discrepancy could be caused by the fact they took into account a preliminary downsampling of the data to a time resolution comparable to the width of the pulse of interest, which results in a loss of signal-to-noise ratio. On the other hand, we have considered the optimal case where the time (or phase) sampling interval is much smaller than the pulse width.

\subsection{Variance of uncorrelated Gaussian noise samples downsampled by a non-integer factor}
\label{subsec:variance_non_integer_downsampling}

\begin{figure}
\centering
\includegraphics[width=0.80\columnwidth]{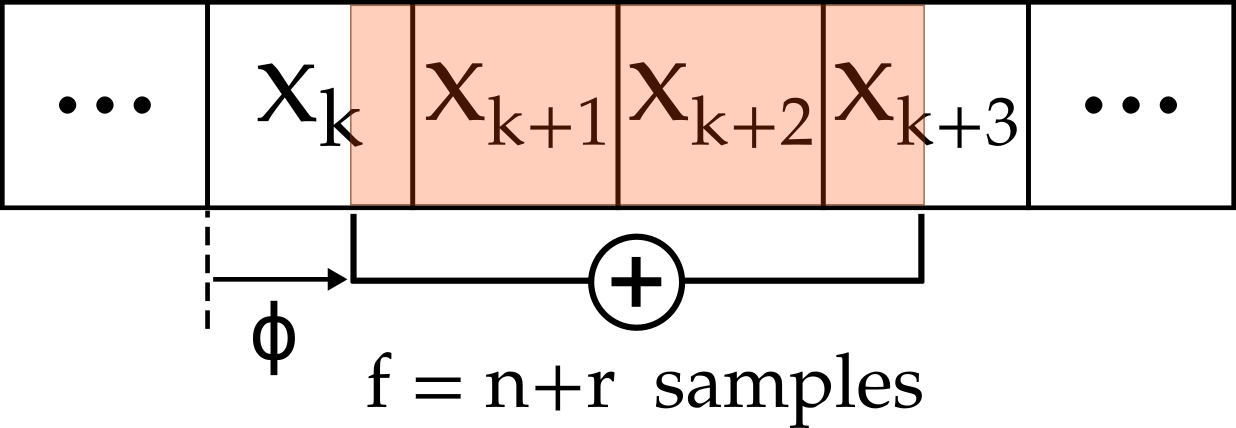}
\caption{An illustration of downsampling a time series by a non-integer factor $f = n + r$, where $n = \floor{f}$ and $0 < r < 1$.}
\label{fig:non_integer_downsampling}
\end{figure}

Here we consider a sequence of time samples $x_i$ made of normally distributed, uncorrelated Gaussian noise, with arbitrary mean and unit variance. The operation of downsampling by a non-integer factor $f = n + r$  where $n = \floor{f}$ and $0 < r < 1$ is illustrated in Fig. \ref{fig:non_integer_downsampling}. A downsampling window gathers and adds $f$ consecutive input samples together into an output sample, before moving forward by $f$ input samples and repeating the process, creating a new time series with a time resolution reduced by a factor $f$. In the addition, samples that only partially overlap with the window are weighted by their overlap fraction. As the window moves forward, the ``initial phase" $\phi$ of the downsamping window relative the start of the first sample with which it overlaps will vary; the variance of the output sample created is a function of $\phi$. Indeed one can distinguish two cases:

\begin{itemize}
    \item Case A: $0 \leq \phi < 1-r $. The window gathers $1 - \phi$ partial sample on the left, $n - 1$ full samples in the middle, and $r+\phi$ partial sample on the right.
    \item Case B: $1-r \leq \phi < 1 $. The window gathers $1 - \phi$ partial sample on the left, $n$ full samples in the middle, and $r+\phi-1$ partial sample on the right.
\end{itemize}
The variance of a \textit{single} output sample $\sigma^2$ as a function of $\phi$ is thus

\begin{equation}
\sigma^2(\phi) = \begin{cases}
(1-\phi)^2 + (n-1) + (r+\phi)^2   &\text{if $0 \leq \phi < 1-r $}\\
(1-\phi)^2 + n + (r+\phi-1)^2     &\text{if $1-r \leq \phi < 1 $}
\end{cases}
\end{equation}
where we used the fact that the input samples are uncorrelated, the general property $\mathrm{Var}[aX] = a^2 \mathrm{Var}[X]$, and that uncorrelated variances add linearly. If $f$ is an irrational number, then over a long input time series $\phi$ will take values that can be considered randomly distributed between 0 and 1. Therefore the \textit{average} variance of output samples $\overline{\sigma^2}$ is

\begin{equation}
    \overline{\sigma^2} = \int_{0}^{1} \sigma^2(\phi) d\phi = f - \frac{1}{3},
\end{equation}
which we have verified numerically by downsampling artificial Gaussian noise. It is interesting to note that if $f$ is a rational number with a small denominator, $\phi$ cycles through a small set of rational values and the above is not quite correct; we neglect that edge case in \textsc{riptide}, where we apply the above equation for all non-integer $f$.

\bsp	
\label{lastpage}
\end{document}